\documentclass[english]{article}
\usepackage{mathpazo}
\usepackage[T1]{fontenc}
\usepackage[latin9]{inputenc}
\usepackage{geometry}
\usepackage{babel}
\usepackage{graphicx}
\usepackage[authoryear]{natbib}
\usepackage[unicode=true]
 {hyperref}

\makeatletter

\providecommand{\tabularnewline}{\\}

\newcommand{\lyxaddress}[1]{
\par {\raggedright #1
\vspace{1.4em}
\noindent\par}
}

\makeatother

\begin{document}

\title{The complex planetary synchronization structure of the solar system}

\author{Nicola Scafetta$^{1,2}$}

\maketitle

\lyxaddress{$^{1}$Active Cavity Radiometer Irradiance Monitor (ACRIM) Lab, Coronado,
CA 92118, USA}

\lyxaddress{$^{2}$Duke University, Durham, NC 27708, USA}
\begin{abstract}
The complex planetary synchronization structure of the solar system,
which since Pythagoras of Samos (ca. 570\textendash{}495 BC) is known
as the \textit{music of the spheres}, is briefly reviewed from the
Renaissance up to contemporary research. Copernicus' heliocentric
model from 1543 suggested that the planets of our solar system form
a kind of mutually ordered and quasi-synchronized system. From 1596
to 1619 Kepler formulated preliminary mathematical relations of approximate
commensurabilities among the planets, which were later reformulated
in the Titius-Bode rule (1766-1772) that successfully predicted the
orbital position of Ceres and Uranus. Following the discovery of the
$\sim$11 yr sunspot cycle, in 1859 Wolf suggested that the observed
solar variability could be approximately synchronized with the orbital
movements of Venus, Earth, Jupiter and Saturn. Modern research have
further confirmed that: (1) the planetary orbital periods can be approximately
deduced from a simple system of resonant frequencies; (2) the solar
system oscillates with a specific set of gravitational frequencies,
and many of them (e.g. within the range between 3 yr and 100 yr) can
be approximately constructed as harmonics of a base period of $\sim$178.38
yr; (3) solar and climate records are also characterized by planetary
harmonics from the monthly to the millennia time scales. This short
review concludes with an emphasis on the contribution of the author's
research on the empirical evidences and physical modeling of both
solar and climate variability based on astronomical harmonics. The
general conclusion is that the solar system works as a resonator characterized
by a specific harmonic planetary structure that synchronizes also
the Sun's activity and the Earth's climate. The special issue \textit{\textquotedbl{}Pattern
in solar variability, their planetary origin and terrestrial impacts\textquotedbl{}}
\citep{Morneretal2013} further develops the ideas about the planetary-solar-terrestrial
interaction with the personal contribution of 10 authors.
\end{abstract}

\section{Introduction}

In 1543 the\textit{ De revolutionibus orbium coelestium (On the Revolutions
of the Heavenly Spheres)} was published. As opposed to Ptolemy's geocentric
model that had been widely accepted since antiquity, \citet{Copernicus}
proposed a heliocentric model for the solar system: the planets, including
the Earth, orbit the Sun and their orbital periods increase with the
planetary distance from the Sun. Copernicus also argued that the planets
form a kind of mutually ordered system. The physical properties of
the planets' orbits, such as their distances from the Sun and their
periods, did not appear randomly distributed. They appeared to obey
a certain law of nature.

A typical synchronization that could be more easily highlighted by
the heliocentric system was, for example, the 8:13 Earth-Venus orbital
resonance. Every 8 years the Earth-Venus orbital configuration approximately
repeats because the Earth revolves 8 times and Venus $\sim$13 times,
as can be easily calculated using their sidereal orbital periods:
$P_{Ea}=365.256$ days and $P_{Ve}=224.701$ days. Figure 1A demonstrates
this orbital regularity by showing the relative positions of Earth
and Venus on January $1^{st}$ from 2012 to 2020. 

However, Venus presents a more subtle and remarkable synchronization
with Earth. The rotation period of Venus on her own axis is 243.021
days (that is almost exactly two-thirds of the Earth's annual period)
and is retrograde. It is easy to calculate that at every inferior
conjunction (that is, every time the Sun, Venus and Earth line up)
the same side of Venus faces Earth \citep{Goldreich,Jelbring2013a};
the Venus-Earth synodic period is 583.924 days and there are 5 inferior
conjunctions in 8 years. In fact, as Figure 1B shows, in one synodic
period Earth revolves 1.59867 times around the Sun and Venus rotates
on its own axis 2.40277 times in the opposite direction: the sum of
the fractional part of the two numbers is almost exactly 1 ($\sim$1.00144).
Thus, not only Earth is almost synchronized with Venus in a 8:13 orbital
resonance and in a 8:5 synodic resonance but, despite the large distance
separating the two planets, Earth seems to have also synchronized
Venus' rotation. It is unlikely that this phenomenon is just a coincidence. 

\begin{center}
\begin{figure}[!t]
\centering{}\includegraphics[width=1\textwidth]{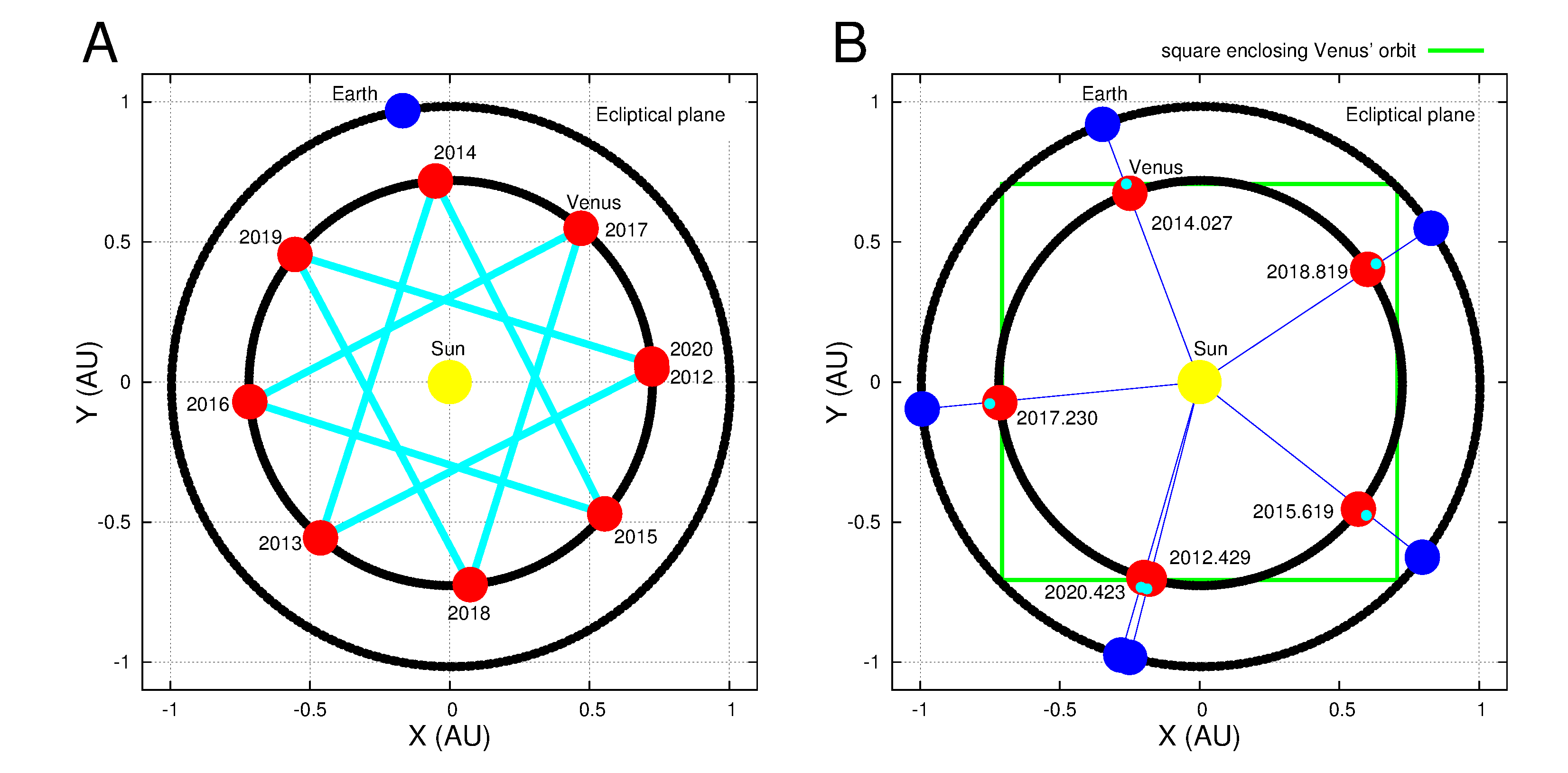}\caption{{[}A{]} Earth and Venus' orbits and their positions on January $1^{st}$
for the years 2012 to 2020 in the Copernicus' heliocentric system.
The figure shows that every $8$ years the Venus-Earth configuration
approximately repeats forming a 8-point star pattern. {[}B{]} Earth-Venus
inferior conjunctions from 2012 to 2020. The figure shows a 5-point
star pattern. Note that at every conjunction the same side of Venus
(represented by a little cyan circle) faces Earth. The orbits and
the coordinates (in astronomical units) of the planets were determined
using the JPL's HORIZONS Ephemeris system. \protect\href{http://ssd.jpl.nasa.gov/horizons.cgi}{http://ssd.jpl.nasa.gov/horizons.cgi}}
\end{figure}

\par\end{center}

Earth sees  always the same face of the Moon. The lunar rotation has
been synchronized with Earth by tidal torque. At least 34 moons of
the solar system (e.g. the Galilean moons of Jupiter) are rotationally
synchronized with their planet (\href{http://en.wikipedia.org/wiki/Synchronous_rotation}{http://en.wikipedia.org/wiki/Synchronous\_{}rotation}).
Also Charon and Pluto are gravitationally locked and keep the same
face toward each other. Mercury rotation period (58.646 days) is exactly
2/3 of its orbital period (87.969 days) \citep{Goldreichb,Jelbring2013a}.
The synchronization of Mercury's rotation with its orbital period
may be due to the combined effect of the strong tidal torque by the
Sun and to Mercury's eccentricity ($\sim0.2$), which implies that
at perihelion Mercury is about 2/3 of its aphelion distance from the
Sun: 0.307 AU versus 0.467 AU. It is also well known that the three
inner moons of Jupiter \textendash{} Ganymede, Europa, and Io \textendash{}
participate in a 1:2:4 orbital resonance. However, the synchronous
rotation of Venus with the Earth's orbit is surprising, given the
large distance between the two planets. In fact, the theoretical tidal
elongation caused by the Earth's gravity on Venus is just a fraction
of millimeter. At the inferior conjunction the tidal elongation caused
by Earth on Venus is maximum and is about $3m_{Ea}R_{Ve}^{4}/2m_{Ve}d_{VE}^{3}=0.035$
mm where $m_{Ea}=1$ and $m_{Ve}=0.815$ are the masses of Earth and
Venus in Earth's mass unit, $R_{Ve}=6051.8$ km is the radius of Venus
and $d_{VE}=41.4\times10^{6}$ km is the average distance between
Earth and Venus at the inferior conjunction. 

Numerous other examples of strong commensurabilities among the planets
of the solar system have been found and some of them will be discussed
in this paper \citep[cf.:][]{Jelbring2013a,Tattersall}. Also the
27.3 days sidereal orbital period of the Moon around Earth appears
well synchronized with the 27.3 days period of the Carrington rotation
of the Sun, as seen from the Earth, which determines a main electromagnetic
oscillation of the heliospheric current sheet in a Parker spiral.
The collective synchronization among all celestial bodies in our solar
system indicates that they interact energetically with each other
and have reached a quasi-synchronized dynamical state. 

Indeed, the bodies of the solar system interact with each other gravitationally
and electromagnetically, and their orbits and rotations are periodic
oscillators. As discovered by Christian Huygens in the 17$^{th}$
century, entrainment or synchronization between coupled oscillators
requires very little energy exchange if enough time is allowed. Huygens
patented the first pendulum clock and first noted that, if hanged
on the same wall, after a while, pendulum clocks synchronize to each
other due to the weak physical coupling induced by small harmonic
vibrations propagating in the wall \citep{Pikovsky}. Note that the
solar system is about 5 billion years old, is not part of a stellar
binary system and in its history should not have experienced particularly
disrupting events such as collisions with other solar systems. Therefore,
a certain degree of harmonic synchronization among its components
should be expected. 

Newtonian mechanics calculates that the theoretical tidal elongation
induced by the gravity of the planets inside the Sun is just a fraction
of millimeter \citep{Scafetta2012c}. Therefore, tidal forcing appears
too small to effect the Sun. However, as discussed above, also the
magnitude of the tidal elongation induced by the Earth's gravity on
Venus is a fraction of millimeter. Thus, if the Earth's gravity or
some other planetary mechanism has synchronized the rotation of Venus
with Earth, the planets could have synchronized the internal dynamics
of the Sun and, therefore, they could be modulating solar activity.
It just seems unlikely that in a solar system where everything looks
more or less synchronized with everything else, only the Sun should
not be synchronized in some complex way with planetary motion. 

Then, the Earth's climate could be modulated by a complex harmonic
forcing made of: (1) lunar tidal oscillations acting mostly in the
ocean; (2) planetary induced solar luminosity and electromagnetic
oscillations modulating mostly the cloud cover and, therefore, the
Earth's albedo; (3) a gravitational synchronization with the moon
and other planets of the solar system modulating, for example, the
Earth's orbital trajectory and its length of the day \citep[cf.][]{Morner2013}. 

From Kepler\textquoteright{}s basic concepts forward through time,
this paper briefly summarizes some of the results that have further
suggested the existence of a complex synchronization structure permeating
the entire solar system whose physical origin is still not fully understood.
A number of empirical studies have shown that a complex synchronized
planetary harmonic order may characterize not only the solar planetary
system but also the Sun's activity and the Earth's climate, fully
confirming Kepler's vision about the existence of a \textit{harmony
of the world.} Preliminary physical mechanisms are being proposed
as well. 

This brief review is not fully comprehensive of all the results. It
simply introduces a general reader to this fascinating issue. The
next sections review general results found in the scientific literature
about: (1) the ordered structure of the planetary system; (2) the
likely planetary origin of the oscillations of the Sun's activity;
(3) the synchronization of the Earth's climate with lunar, planetary
and solar harmonics. 

\begin{center}
\begin{figure}[!t]
\centering{}\includegraphics[width=0.6\paperwidth]{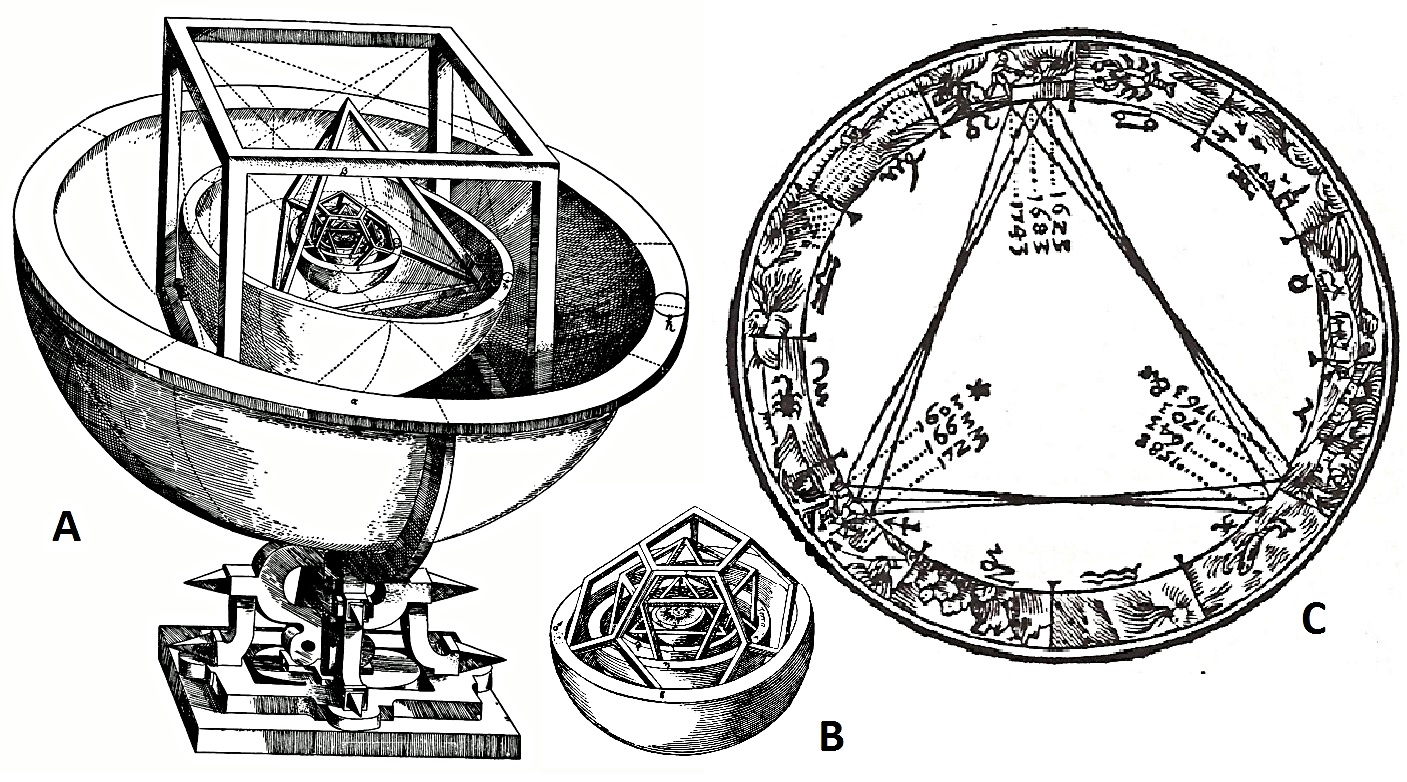}\caption{{[}A{]} Encapsulated platonic solid model of the solar planetary system
\citep{Kepler1596}. {[}B{]} Detailed view of the inner sphere. {[}C{]}
A series of Great Conjunction of Jupiter and Saturn from 1583 to 1723
by \citet{Kepler1606}. The figure demonstrates that every $\sim$60
years the Jupier-Saturn configuration approximately repeats. Every
$\sim$20 years a Jupier-Saturn conjunction occurs. (Figures are adapted
from \protect\href{http://en.wikipedia.org}{http://en.wikipedia.org}).}
\end{figure}

\par\end{center}

\section{Kepler's vision of a cosmographic mystery}

About half century after Copernicus, Kepler corrected and extended
the heliocentric model. Kepler discovered that: (1) the orbit of every
planet is an ellipse (instead of Copernicus' perfect cycles) with
the Sun at one of the two foci (instead of being in the center of
the cycle); (2) a line joining a planet and the Sun sweeps out equal
areas during equal intervals of time; (3) the square of the orbital
period of a planet is proportional to the cube of the semi-major axis
of its orbit. If the orbital period, $T$, is measured in years and
the semi-major axis, $a$, is measured in astronomical units (AU,
the average Sun-Earth distance), Kepler's third law takes the simple
form of $T^{2}=a^{3}$. The first two laws were published in 1609
\citep{Kepler1609}, while the third law was published in 1619 \citep{Kepler1619}.
Kepler's three laws of planetary motion were later formally demonstrated
by \citet{Newton} using calculus and his law of universal gravitation
stating that a planet is attracted by the Sun with a force directly
proportional to the product of the two masses and inversely proportional
to the square of the sun-planet distance.

However, Kepler's did more than just proposing his three laws of planetary
motion. Since the publication of the \textit{Mysterium Cosmographicum
}(The Cosmographic Mystery) \citet{Kepler1596} noted the existence
of a \textit{marvelous proportion of the celestial spheres} referring
on the \textit{number, magnitude, and periodic motions of the heavens}.
Kepler proposed that the distance relationships between the six planets
known at that time (Mercury, Venus, Earth, Mars, Jupiter and Saturn)
could be understood in terms of the five Platonic solids, enclosed
within each other with the outer solid being a sphere that represented
the orbit of Saturn. See Figures 2A and 2B. 

Some of these geometrical relations are easy to notice. For example,
the ratio between the Earth's orbital radius ($a=1$ AU) and Venus'
orbital radius ($a=0.72$ AU) is approximately equal to the ratio
between the diagonal and the side of a square ($\sqrt{2}\approx1.41$).
Thus, Venus' orbit is approximately enclosed within a square enclosed
within the Earth's orbit: see Figure 1B. Analogously, the ratio between
Saturn's orbital radius ($a=9.6$ AU) and Jupiter's orbital radius
($a=5.2$ AU) is approximately equivalent to the ratio between the
diagonal and the side of a cube ($\sqrt{3}\approx1.73$). Thus, Jupiter's
orbit is approximately enclosed within a cube enclosed within Saturn's
orbital sphere: see Figure 2A.

Kepler also highlighted the existence of a 5:2 Jupiter-Saturn resonance
that was, however, well known since antiquity \citep{Masar,Temple}:
every $\sim$60 years the Jupiter-Saturn configuration approximately
repeats because Jupiter revolves $\sim$5 times and Saturn $\sim$2
times. Figure 2C shows the original Kepler's diagram of the great
conjunctions of Saturn and Jupiter, which occur every $\sim$20 yr,
from 1583 to 1723. Every three conjunctions (a \textit{trigon}) Jupiter
and Saturn meet approximately at the same location of the zodiac,
which happens every $\sim$60 yr. The \textit{trigon} slightly rotates
and the configuration repeats every 800-1000 years. 

The discovery of a geometrical relationship among the semi-major axes
of the planets and the relationship between the planets' orbital semi-major
axis and their orbital period (the $3^{rd}$ law of planetary motion)
convinced \citet{Kepler1619} that the planetary orbits are mutually
synchronized as though the solar system formed a kind of \textit{celestial
choir}. The great advantage of the heliocentric model was mostly to
make it far easier to see this ordered structure. 

Kepler also conjectured that celestial harmonics could permeate the
entire solar system including the Earth's climate \citep{Kepler1601,Kepler1606,Kepler1619}.
However, modern physics suggests that for the planets to modulate
the Earth's climate, they first have to synchronize the Sun's activity.
In fact, the Sun is the most likely place where the weak planetary
tidal harmonics could be energetically amplified by a large factor.
This issue will be discussed in Sects. 7 and 8.

\section{The planetary rhythm of the Titius-Bode rule}

\citet{Titius} and later \citet{Bode} noted that the semi-major
axes $a_{n}$ of the planets of the solar system are function of the
planetary sequence number $n$. Adding 4 to the series 0, 3, 6, 12,
24, 48, 96, 192 \& 384 and dividing the result by 10 gives a series
that approximately reproduces the semi-major axis length of the planets
in astronomical units (1 AU = Sun-Earth average distance). The Titius-Bode
rule for the orbital semi-major axis length, $a_{n}$, is a power-law
equation that can be written as 

\begin{equation}
a_{n}=0.4+0.3*2^{n}
\end{equation}
with $n=-\infty$, 0, 1, 2, 3, 4, 5, 6, 7, where $n=-\infty$ refers
to Mercury, $n=0$ to Venus, $n=1$ to Earth, etc. As Table 1 shows,
the Titius-Bode empirical rule successfully predicts the orbital semi-major
axis length for all the planets and dwarf planets, except for Neptune. 

\begin{table}
\centering{}%
\begin{tabular}{|c|c|c|c|c|}
\hline 
planet & n & Titius-Bode rule & observations & percent error\tabularnewline
 &  & $a_{n}$ (AU) & $a$ (AU) & \tabularnewline
\hline 
\hline 
Mercury & $-\infty$ & 0.40 & 0.387 & (3.3\%)\tabularnewline
\hline 
Venus & 0 & 0.70 & 0.723 & (3.18\%)\tabularnewline
\hline 
Earth & 1 & 1.00 & 1.00 & (0\%)\tabularnewline
\hline 
Mars & 2 & 1.60 & 1.524 & (5.0\%)\tabularnewline
\hline 
Ceres & 3 & 2.80 & 2.77 & (1.1\%)\tabularnewline
\hline 
Jupiter & 4 & 5.20 & 5.204 & (0.1\%)\tabularnewline
\hline 
Saturn & 5 & 10.00 & 9.582 & (4.4\%)\tabularnewline
\hline 
Uranus & 6 & 19.60 & 19.201 & (2.1\%)\tabularnewline
\hline 
Neptune & ? & ? & 30.047 & ?\tabularnewline
\hline 
Pluto & 7 & 38.80 & 39.482 & (1.7\%)\tabularnewline
\hline 
\end{tabular}\caption{Predictions of the Titius-Bode rule against the observations. The
semi-major axes $a$ are measured in Astronomical Units. The observed
semi-major axes are from \protect\href{http://nssdc.gsfc.nasa.gov/planetary/factsheet/}{http://nssdc.gsfc.nasa.gov/planetary/factsheet/}}
\end{table}

When Titius-Bode rule was proposed (1766-1772) the dwarf planet Ceres
(in the asteroid belt) and the Jovian planet Uranus were unknown.
Indeed, the idea that undiscovered planets could exist between the
orbits of Mars and Jupiter and beyond Saturn was strongly suggested
by Bode in 1772. The curious gap separating Mars and Jupiter was,
however, already noted by Kepler. 

The astronomers looked for new planets taking into account the predictions
of the Titius-Bode rule. In 1781 Herschel \citep{Dreyer} discovered
Uranus, and in 1801 \citet{Piazzi} discovered the dwarf planet Ceres.
Both Ceres and Uranus fit the predictions of the Titius-Bode rule
relatively well. 

In the early 19$^{th}$ century, following Herschel and Piazzi\textquoteright{}s
discoveries, the Titius-Bode rule became widely accepted as a \textquotedblleft{}law\textquotedblright{}
of nature. However, the discovery of Neptune in 1846 created a severe
problem because its semi-major axis length $a_{Ne}=30.047$ AU does
not satisfy the Titius-Bode prediction for $n=7$, $a_{7}=38.80$
AU. The discovery of Pluto in 1930 confounded the issue still further.
In fact, Pluto's semi-major axis length, $a_{pl}=39.482$ AU, would
be inconsistent with the Titius-Bode's rule unless Pluto is given
the position $n=7$ that the rule had predicted for Neptune. See Table
1.

The Titius\textendash{}Bode rule is clearly imperfect or incomplete
and no rigorous theoretical explanation of it still exists. However,
it is unlikely that the relationship among the planets of the solar
system that it approximately models is purely coincidental. Very likely
any stable planetary system may satisfy a Titius\textendash{}Bode-type
relationship due to a combination of orbital resonance and shortage
of degrees of freedom. \citet{Dubrullea,Dubrulleb} have shown that
Titius\textendash{}Bode-type rules could be a consequence of collapsing-cloud
models of planetary systems possessing two symmetries: rotational
invariance and scale invariance.

\section{The Asteroid Belt ``mirror'' symmetry rule}

Following the discovery of Ceres in 1801, numerous asteroids were
discovered at approximately the same orbital distance. The region
in which these asteroids were found lies between Mars and Jupiter
and it is known as the Asteroid Belt. No planet could form in this
region because of the gravitational perturbations of Jupiter that
has prevented the accretion of the asteroids into a small planet.
Ceres, with its spherical shape of $\sim$500 Km radius, is just the
largest asteroid and the only dwarf planet in the inner solar system.

A curious mathematical relationship linking the four terrestrial inner
planets (Mercury, Venus, Earth and Mars) and the four giant gaseous
outer planets (Jupiter, Saturn, Uranus and Neptune) exists \citep{Geddes}.
The semi-major axes of these eight planets appear to \textit{reflect}
about the asteroid belt. This mirror symmetry associates Mercury with
Neptune, Venus with Uranus, Earth with Saturn and Mars with Jupiter.
\citet{Geddes} found that the mutual relations among the planets
could all be approximately given as relations between the mean frequency
notes in an octave: $b=2\exp(1/8)$.

For example, using the semi-major axis lengths reported in Table 1
for the eight planets and labeling these distances with the first
two letters of the planet's name, it is easy to obtain:

\begin{equation}
\begin{array}{ccc}
Me\times Ne & = & 1.214\cdot Ea\times Sa\\
Ve\times Ur & = & 1.194\cdot Me\times Ne\\
Ea\times Sa & = & 1.208\cdot Ma\times Ju
\end{array}
\end{equation}
where we have $b^{2}\approx1.19$, and

\begin{equation}
\begin{array}{ccc}
Ve\times Ma & = & 2.847\cdot Me\times Ea\\
Sa\times Ne & = & 2.881\cdot Ju\times Ur
\end{array}
\end{equation}
where we have $b^{12}\approx2.83$. Combining the equations yields:

\begin{equation}
\frac{Me\times Ne}{Ea\times Sa}\approx\frac{Ve\times Ur}{Me\times Ne}\approx\frac{Ea\times Sa}{Ma\times Ju}
\end{equation}
and

\begin{equation}
\frac{Me\times Ea}{Ve\times Ma}\approx\frac{Ju\times Ur}{Sa\times Ne}.
\end{equation}
These relations relate the four inner and the four outer planets of
the solar system. Even if Geddes and King-Hele rule is not perfect,
it does suggest the existence of a specific ordered structure in the
planetary system where the asteroid belt region plays a kind of mirroring
boundary condition between the inner and outer regions of the solar
system.

\citet{Geddes} concluded: \textit{``The significance of the many
near-equalities is very difficult to assess. The hard-boiled may dismiss
them as mere playing with numbers; but those with eyes to see and
ears to hear may find traces of something far more deeply interfused
in the fact that the average interval between the musical notes emerges
as the only numerical constant required \textendash{} a result that
would surely have pleased Kepler.''}

\section{The matrix of planetary resonances}

\citet{Molchanov1968,Molchanov1969a} showed that the periods of the
planets could be approximately predicted with a set of simple linear
equations based on integer coefficients describing the mutual planetary
resonances. Molchanov's system is reported below: 

\begin{equation}
\left(\begin{array}{ccccccccc}
1 & -1 & -2 & -1 & 0 & 0 & 0 & 0 & 0\\
0 & 1 & 0 & -3 & 0 & -1 & 0 & 0 & 0\\
0 & 0 & 1 & -2 & 1 & -1 & 1 & 0 & 0\\
0 & 0 & 0 & 1 & -6 & 0 & -2 & 0 & 0\\
0 & 0 & 0 & 0 & 2 & -5 & 0 & 0 & 0\\
0 & 0 & 0 & 0 & 1 & 0 & -7 & 0 & 0\\
0 & 0 & 0 & 0 & 0 & 0 & 1 & -2 & 0\\
0 & 0 & 0 & 0 & 0 & 0 & 1 & 0 & -3
\end{array}\right)\left(\begin{array}{c}
\omega_{Me}\\
\omega_{Ve}\\
\omega_{Ea}\\
\omega_{Ma}\\
\omega_{Ju}\\
\omega_{Sa}\\
\omega_{Ur}\\
\omega_{Ne}\\
\omega_{Pl}
\end{array}\right)=\left(\begin{array}{c}
0\\
0\\
0\\
0\\
0\\
0\\
0\\
0
\end{array}\right),\label{eq:m1}
\end{equation}
where $\omega=T^{-1}$ is the orbital frequency corresponding to the
planetary period $T$. By imposing $\omega_{Ea}^{-1}=T_{Ea}=1$ yr
the system (Eqs. \ref{eq:m1}) predicts the following orbital periods: 

\begin{equation}
\begin{array}{cclcccc}
period &  &  &  & calculated & observed & error\\
T_{Me} & = & 2484/10332 & = & 0.240 & 0.241 & \left(0.4\%\right)\\
T_{Ve} & = & 2484/4044 & = & 0.614 & 0.615 & \left(0.2\%\right)\\
T_{Ea} & = & 1 & = & 1.000 & 1.000 & \left(0.0\%\right)\\
T_{Ma} & = & 2484/1320 & = & 1.880 & 1.880 & \left(0.0\%\right)\\
T_{Ju} & = & 2484/210 & = & 11.83 & 11.86 & \left(0.3\%\right)\\
T_{Sa} & = & 2484/84 & = & 29.57 & 29.46 & \left(0.4\%\right)\\
T_{Ur} & = & 2484/30 & = & 82.80 & 84.01 & \left(1.4\%\right)\\
T_{Ne} & = & 2484/15 & = & 165.6 & 164.8 & \left(0.5\%\right)\\
T_{Pl} & = & 2484/10 & = & 248.4 & 248.1 & \left(0.1\%\right)
\end{array}
\end{equation}
where the last column gives in years the observed orbital periods
of the planets. The absolute percent divergence between the predicted
and observed orbital periods is given in parenthesis.

Using simple linear algebra, the system (Eqs. \ref{eq:m1}) can also
be used to find alternative resonance relations. For example, summing
the first two rows gives the following relation between Mercury, Earth,
Mars and Jupiter: $\omega_{Me}-2\omega_{Ea}-4\omega_{Ma}-\omega_{Sa}=0$.

\citet{Molchanov1968} showed that analogous tables of integers work
also for describing planetary satellite systems such as the moon system
of Jupiter and Saturn. The provided physical explanation was that
resonant structure in a gravitationally interacting oscillating system
could be inevitable under the action of dissipative perturbations
of mutually comparable size. However, \citet{Molchanov1969a} noted
that alternative resonance relations yielding slightly different results
could also be formulated. Nevertheless, even if it is the case that
the system (Eqs. \ref{eq:m1}) is neither unique nor perfectly descriptive
of the orbital characteristics of the planets of the solar system,
it does suggest that the planets are mutually synchronized. \citet{Molchanov1969b}
quantitatively evaluated that the probability of formation of a given
resonant structure by chance is not very likely: the probability that
the resonant structure of the solar system could emerge as a random
chance was calculated to be less than $p=10^{-10}$.

\section{The gravitational harmonics of the solar system}

The simplest way to determine whether the solar system is characterized
by a harmonic order is to study its natural frequencies and find out
whether they obey some general rule. The main set of frequencies that
characterize the solar planetary system can be found by studying the
power spectra of physical measures that are comprehensive of the motion
of all planets such as the functions describing the dynamics of the
Sun relative to the center-of-mass of the solar system. In fact, the
Sun is wobbling around the center-of-mass of the solar system following
a very complex trajectory due to the gravitational attraction of all
planets. Figure 3 shows the wobbling of the Sun during specific periods. 

\begin{figure}[!t]
\begin{centering}
\includegraphics[width=1\textwidth]{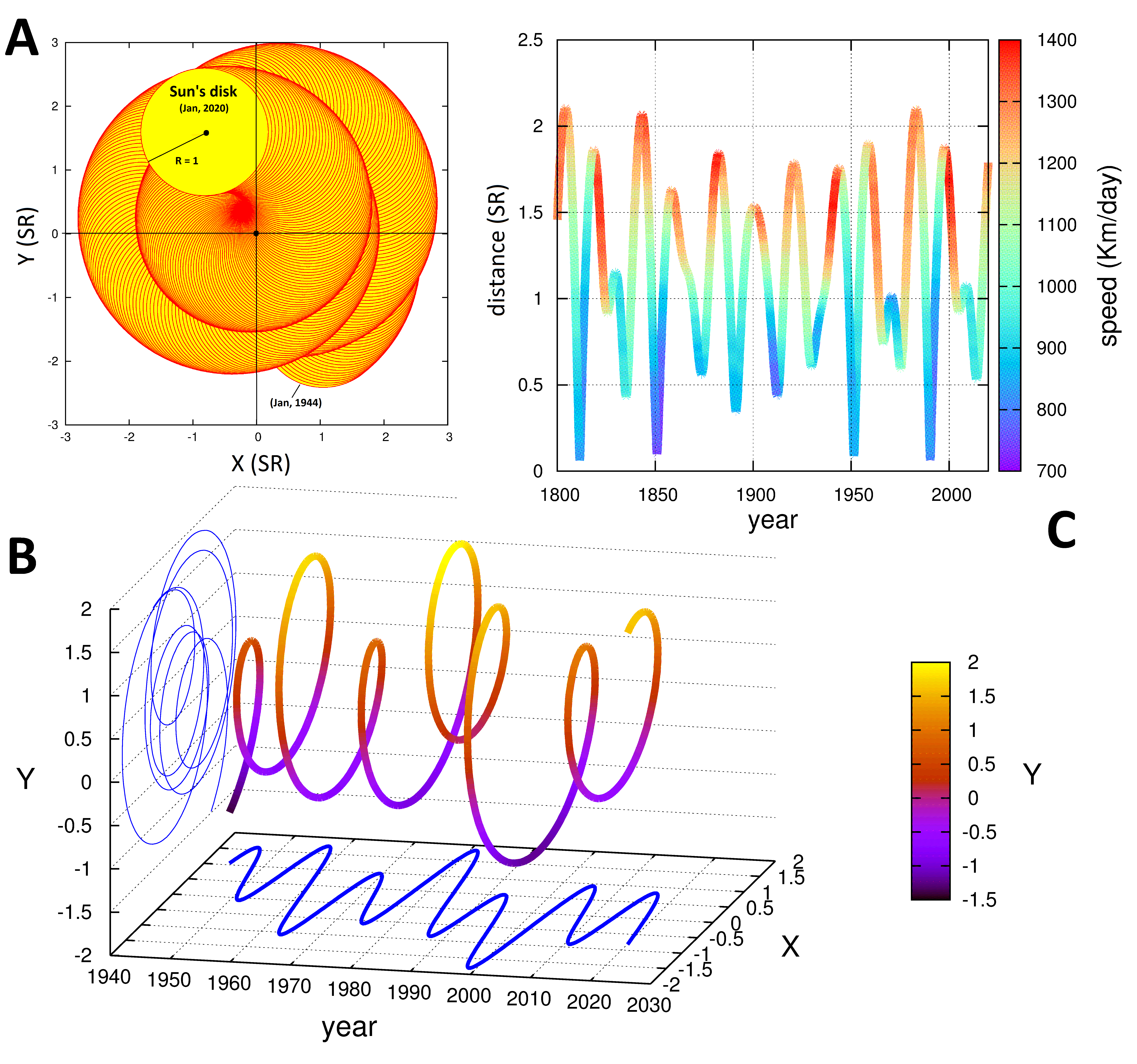}\caption{The wobbling of the Sun relative to the center-of-mass of the solar
system. {[}A{]} Monthly scale movement of the Sun from 1944 to 2020
as seen from the Z-axis perpendicular to the ecliptic. The Sun is
represented by a moving yellow disk with a red circumference \citep[cf.][]{Ebner}.
{[}B{]} The trajectory of the center of the Sun from 1944 to 2020.
{[}C{]} The distance and the speed of the Sun from 1800 to 2020: note
the evident $\sim$20 yr oscillation, and the less evident $\sim$60
and $\sim$170 yr oscillation. The Sun's coordinates are estimated
using the JPL's HORIZONS Ephemeris system. The coordinates are expressed
in solar radius (SR) units.}

\par\end{centering}

\end{figure}

Several functions of the coordinates of the Sun relative to the center-of-mass
of the solar system can be chosen such as the distance, the speed,
the angular momentum etc. \citep[e.g.:][]{Jose,Bucha}. However, simple
mathematical theorems establish that generic functions of the orbits
of the planets must by necessity share a common set of planetary frequencies.
Only the amplitudes associated to each harmonic are expected to depend
on the specific chosen observable. Thus, unless one is interested
in a given observable for a specific purpose, any general function
of the orbits of the planets should suffice to determine the main
harmonic set describing the planetary motion of the solar system as
a whole. 

Herein I extend the frequency analysis of the Sun's motion made in
\citet{Bucha} and \citet{Scafetta2010}. The JPL's HORIZONS Ephemeris
system is used to calculate the speed of the Sun relative to the center-of-mass
of the solar system from BC 8002 Dec 12, to AD 9001 Apr 24 (100 day
steps). Power spectra are evaluated using the periodogram and the
maximum entropy method \citep{Press}. 

\begin{figure}[!t]
\centering{}\includegraphics[width=1\textwidth]{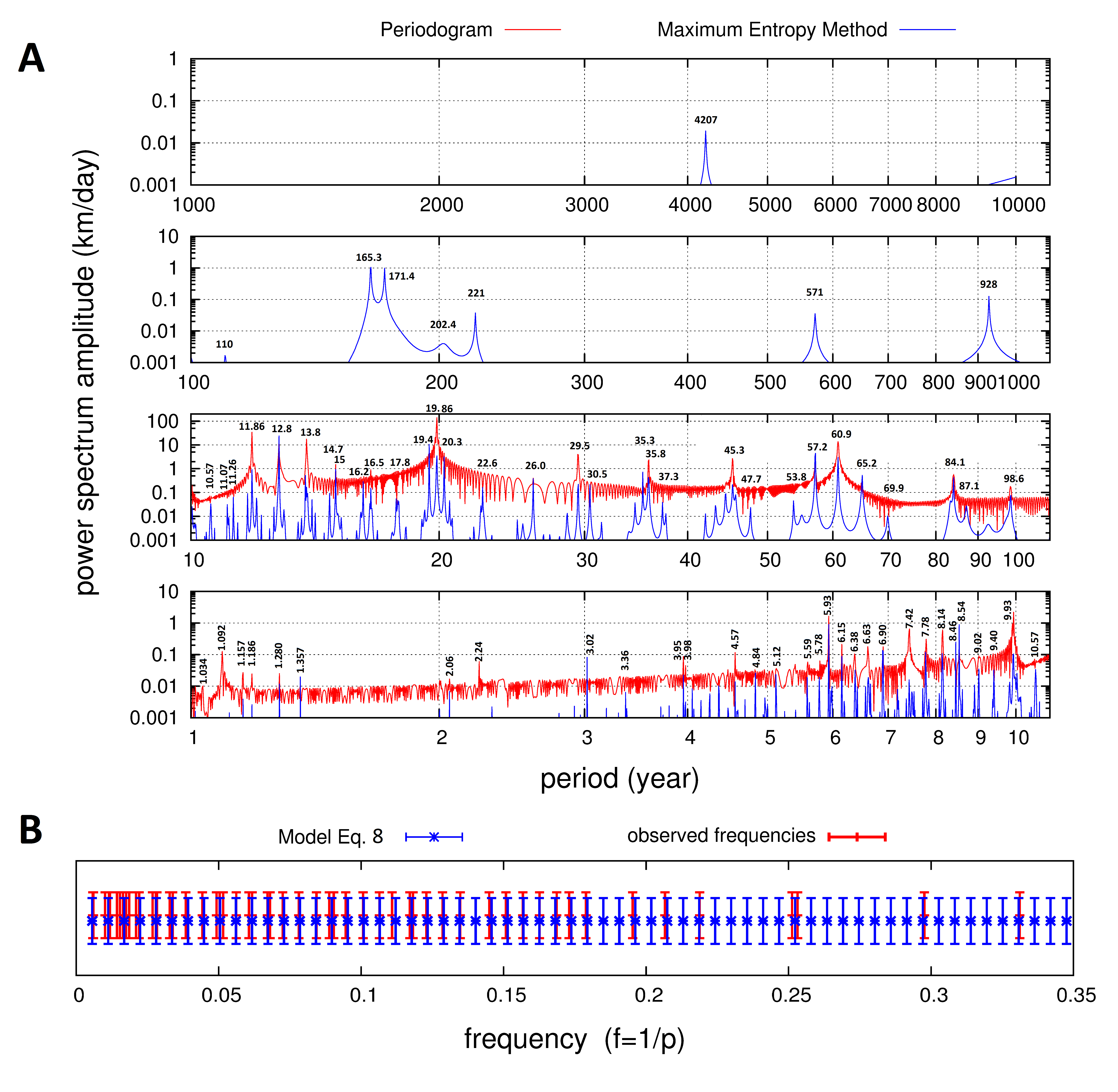}\caption{{[}A{]} Periodogram (red) and the maximum entropy method (blue) of
the speed of the Sun relative to the center-of-mass of the solar system
from BC 8002-Dec-12, to AD 9001-Apr-24. For periods larger than 200
years the periodogram becomes unstable and it is not shown. {[}B{]}
Comparison between the frequencies observed and listed in {[}A{]}
in the range 3 to 200 yr (red) and the frequency predictions of the
resonance model Eq. \ref{eq:bu} (blue). Note the good spectral coherence
of the harmonic model with the observed frequencies.}
\end{figure}

Figure 4A depicts the result and highlights the main planetary frequencies
of the solar system. Slightly different values may be found using
different observables and sub-intervals of the analyzed period because
of statistical variability and because of the relative amplitude of
the frequencies' change with the specific function of the planets'
orbits that are chosen for the analysis. An estimate of the statistical
theoretical error associated to each measured frequency could be obtained
using the Nyquist theorem of the Fourier analysis and it is given
by $\nabla f=\pm1/2L$ where $L=17003$ yr is the length of the analyzed
time sequence. Thus, if $P_{0}$ is the central estimate of a period,
its range is given by $P\approx P_{0}\pm P_{0}^{2}/2L$ \citep[cf. ][]{Tan}. 

Several spectral peaks can be recognized such as the $\sim$1.092
yr period of the Earth-Jupiter conjunctions; the $\sim$9.93 and $\sim$19.86
yr periods of the Jupiter-Saturn spring (half synodic) and synodic
cycles, respectively; the $\sim$11.86, $\sim$29.5,$\sim$84 and
$\sim$165 yr of the orbital period of Jupiter, Saturn, Uranus and
Neptune, respectively; the $\sim$61 yr cycle of the tidal beat between
Jupiter and Saturn; the periods corresponding to the synodic cycle
between Jupiter and Neptune ($\sim$12.8 yr), Jupiter and Uranus ($\sim$13.8
yr), Saturn and Neptune ($\sim$35.8 yr), Saturn and Uranus ($\sim$45.3)
and Uranus and Neptune ($\sim$171.4 yr) and many other cycles including
the spring (half synodic) periods. Additional spectra peaks at $\sim$200-220,
$\sim$571, $\sim$928 and $\sim$4200 yr are also observed. Clustered
frequencies are typically observed. For example, the ranges 42-48
yr, 54-70 yr, 82-100 yr (Gleissberg cycle), 150-230 yr (Suess - de
Vries cycle) are clearly observed in Figure 4 and are also found among
typical main solar activity and aurora cycle frequencies \citep{Ogurtsov,ScafettaWillson2013a}.
The sub-annual planetary harmonics together with their spectral coherence
with satellite total solar irradiance records and other solar records
are discussed in \citet{ScafettaWillson2013b,ScafettaWillson2013c},
and are not reported here.

The curious fact is that the numerous spectral peaks observed in the
solar motion do not seem to be randomly distributed. They could be
approximately reproduced using a simple empirical harmonic formula
of the type \citep{Jakubcova}

\begin{equation}
p_{i}=178.38/i\quad yr,\qquad i=1,2,3,\ldots,\label{eq:bu}
\end{equation}
where the basic period of $\sim$178.38 yr is approximately the period
that \citet{Jose} found in the Sun's motion and in the sunspot record
\citep[cf.:][]{Charvatova2013}. A comparison between the observed
frequencies and the prediction of the resonance model, Eq. \ref{eq:bu},
is shown in Figure 4B. 

Although Eq. \ref{eq:bu} is not perfect, nor all the modeled frequencies
clearly are observed in Figure 4A, the good agreement observed between
most of the observed periods and the harmonic model predictions suggests
that the solar systems is characterized by a complex synchronized
harmonic structure. \citet{Jakubcova} also noted that several spectral
peaks in the solar motion approximately correspond with the periods
of various solar and terrestrial phenomena suggesting that the Sun
itself, and the Earth's climate, could be modulated by the same planetary
harmonics: see also \citet{Charvatova2013}. This issue is further
discussed below.

\section{The planetary synchronization and modulation of the $\sim$11 yr
solar cycle}

In the $19^{th}$ century, solar scientists discovered that sunspot
activity is modulated by a quasi 11 yr oscillation called the Schwabe
cycle. In a letter to Mr. Carrington, \citet{Wolf1859} proposed that
the observed solar oscillation could be caused by the combined influence
of Venus, Earth, Jupiter and Saturn upon the Sun. 

The planetary theory of solar variation is today not favored among
solar scientists because, according to Newtonian physics, the planets
appear too far from the Sun to modulate its activity, for example
by gravitationally forcing the Sun's tachocline \citep{Callebaut}.
The planets could modulate solar activity only if a mechanism that
strongly amplifies their gravitational and/or electromagnetic influence
on the Sun exists. \citet{Scafetta2012c} showed that a strong amplification
mechanism could be derived from the mass-luminosity relation: the
gravitational energy dissipated by planetary tides in the Sun was
proposed to modulate the nuclear fusion rate yielding a variable solar
luminosity production. It was calculated that the proposed mechanism
could yield a $4\times10^{6}$ energetic amplification of the tidal
signal. The derived oscillating luminosity signal could be sufficiently
strong to modulate the Sun's tachocline and convective zone \citep[cf.:][]{Abreu,Morner2013,Solheim}.
Electromagnetic interactions between the planets and the Sun via Parker\textquoteright{}s
spiral magnetic field of the heliosphere, which could be modulated
by functions related to the wobbling dynamics of the Sun such as its
speed, jerk etc., could also be possible in principle. Evidences for
a planet-induced stellar activity has been also observed in other
stars \citep[e.g.: ][]{Scharf,Shkolnik2003,Shkolnik2005}.

It is important to stress that the contemporary view of solar science
is that solar magnetic and radiant variability is intrinsically chaotic,
driven by internal solar dynamics alone and characterized by hydromagnetic
solar dynamo models \citep{Tobias}. However, as also admitted by
solar physicists \citep[e.g.: ][]{deJager,Callebaut}, present hydromagnetic
solar dynamo models, although able to generically describe the periodicities
and the polarity reversal of solar activity, are not yet able to quantitatively
explain the observed solar oscillations. For example, they do not
explain why the Sun should present an $\sim$11 yr sunspot cycle and
a $\sim$22 yr Hale solar magnetic cycle. Solar dynamo models are
able to reproduce an $~$11 yr oscillation only by adopting specific
values for their free parameters \citep{Jiang}. These dynamo models
able to explain also the other solar oscillations observed at multiple
scales such as the 50-140 yr Gleissberg cycle, the 160-260 yr Suess
- de Vries cycle, the millennial solar cycles etc. \citep[cf.][]{Ogurtsov},
nor are they able to explain the phases of these cycles. Thus, the
present solar dynamo theories appear to be incomplete. They cannot
predict solar activity and they have not been able to explain the
complex variability of the solar dynamo including the emergence of
the $\sim$11 yr oscillation. Some mechanism, which is still missed
in the solar dynamo models, is needed to \textit{inform} the Sun that
it needs to oscillate at the observed specific frequencies and at
the observed specific phases. 

However, since \citet{Wolf1859} several studies have highlighted
that the complex variability of the solar dynamo appears to be approximately
synchronized to planetary harmonics at multiple timescales spanning
from a few days to millennia \citep[e.g.: ][ and others]{Abreu,Bigg,Brown,Charvatova,Charvatova2013,Fairbridge,Hung,Jakubcova,Jose,Scafetta2010,Scafetta(2012a),Scafetta2012b,Scafetta2012c,Salvador,Scafetta2012d,Scafetta2013b,ScafettaWillson2013b,ScafettaWillson2013a,ScafettaWillson2013c,Sharp,Solheim,Tan,Wilson,Wolff}.
\citet{Hung} also reports that 25 of the 38 largest known solar flares
were observed to start when one or more tide-producing planets (Mercury,
Venus, Earth, and Jupiter) were either nearly above the event positions
(less than 10 deg. longitude) or at the opposing side of the Sun.

As \citet{Wolf1859} proposed, the $\sim$11 yr solar cycle could
be produced by a combined influence of Venus, Earth, Jupiter and Saturn.
There are two main motivations for this proposal: 

\begin{figure}[!t]
\includegraphics[angle=-90,width=1\textwidth]{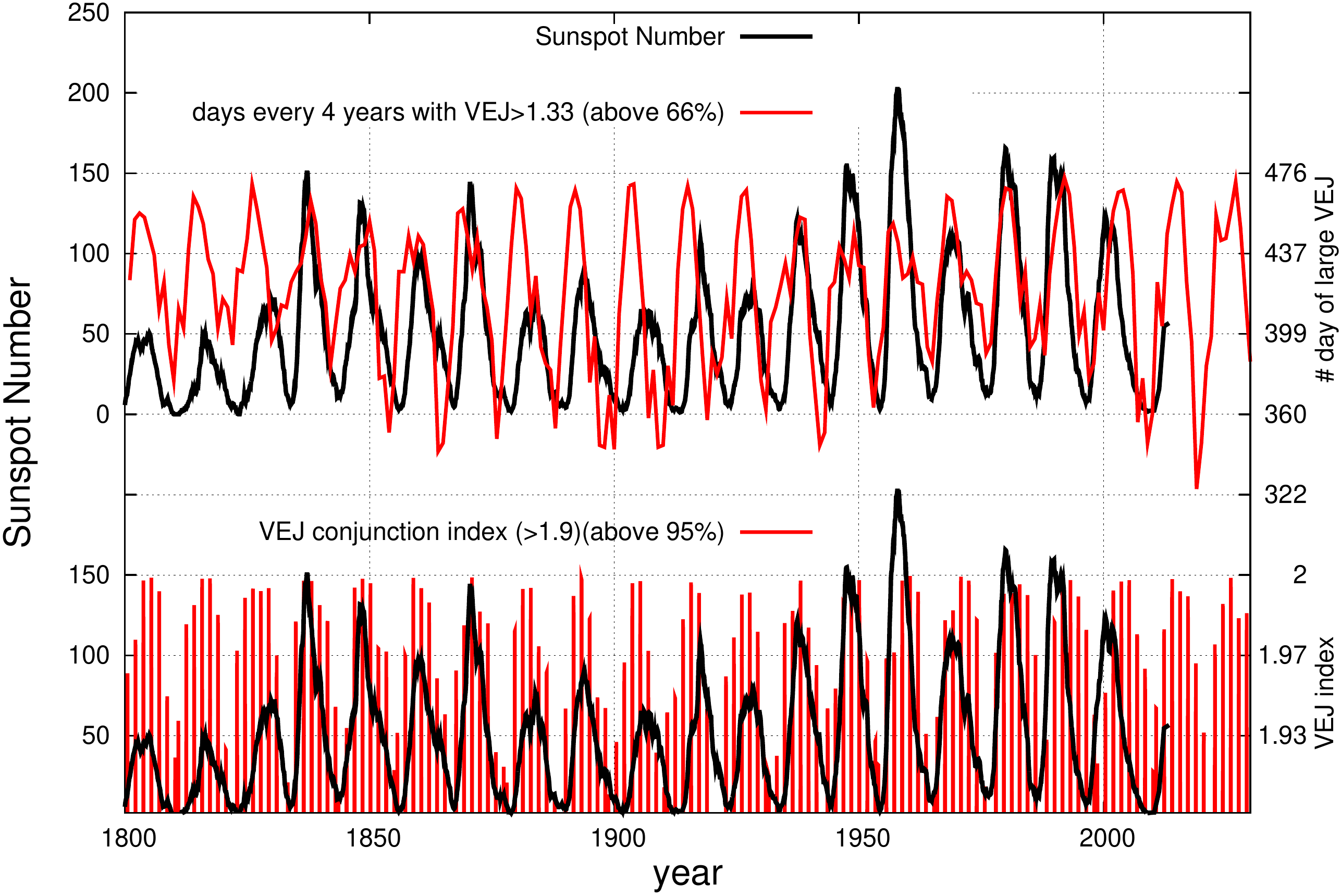}\caption{(Top) the sunspot number record (black) is compared against the number
of days (every four years) (red) when the alignment index $I_{VEJ}>66\%$.
(Bottom) the sunspot number record (black) is compared against the
most aligned days with $I_{VEJ}>95\%$ (red). For details see \citet{Hung}
and \citet{Scafetta2012c}.}

\end{figure}

(1) The first model relating the 11-year solar cycle to the configuration
of Venus, Earth and Jupiter was proposed by \citet{Bendandi}: later
\citet{Bollinger}, \citet{Hung} and others developed equivalent
models. It was observed that Venus, Earth and Jupiter are the three
major tidal planets \citep[e.g.][]{Scafetta2012c}. By taking into
account the combined alignment of Venus, Earth and Jupiter it is easy
to demonstrate that the gravitational configuration of the three planets
repeats every:

\begin{equation}
P_{VEJ}=\left(\frac{3}{P_{Ve}}-\frac{5}{P_{Ea}}+\frac{2}{P_{Ju}}\right)^{-1}=22.14\: yr
\end{equation}
where $P_{Ve}=224.701$ days, $P_{Ea}=365.256$ days and $P_{Ju}=4332.589$
days are the sidereal orbital periods of Venus, Earth and Jupiter,
respectively \citep{Scafetta2012c}. The 22.14 yr period is very close
to the $\sim$22 yr Hale solar magnetic cycle. Moreover, because the
configurations Ea\textendash{}Ve\textendash{}Sun\textendash{}Ju and
Sun\textendash{}Ve\textendash{}Ea\textendash{}Ju are equivalent about
the tidal potential, the tidal cycle presents a recurrence of half
of the above value, that is, a period of $11.07$ yr. This is the
average solar cycle length observed since 1750 \citep[e.g.][]{Scafetta2012b}.
Figure 5 shows that a measure based on the most aligned days among
Venus, Earth and Jupiter is well correlated, in phase and frequency,
with the $\sim$11 yr solar cycle: for details about the Venus-Earth-Jupiter
11.07 yr cycle see \citet[Bendandi's model]{Battistini}, \citet{Bollinger},
\citet{Hung}, \citet{Scafetta2012c}, \citet{Salvador}, \citet{Wilson}
and \citet{Tattersall}. 

\begin{figure}[!t]
\includegraphics[width=1\textwidth]{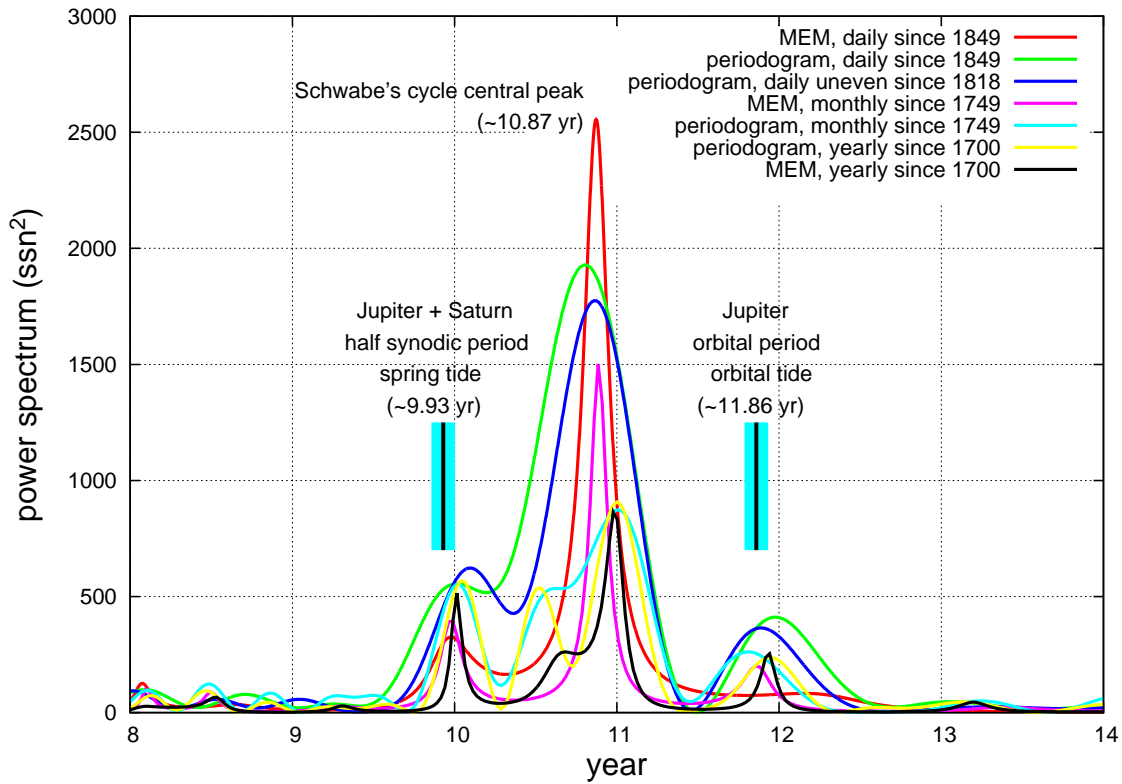}\caption{The three spectral peak structure of the Schwabe's $\sim$11 yr sunspot
cycle as resolved by power spectra estimated using the Maximum Entropy
Method (MEM) and the periodogram \citep{Press}. The two side peaks
at $\sim$9.93 yr and $\sim$11.86 yr correspond to the periods of
Jupiter and Saturn's spring tide and of Jupiter's orbital tide on
the Sun, respectively \citep[cf.: ][]{Scafetta2012b,Scafetta2012c,Solheim}.
Daily, monthly and yearly resolved sunspot number records are used
covering periods from 1700 to 2013: \protect\href{http://sidc.oma.be/sunspot-data/}{http://sidc.oma.be/sunspot-data/}}
\end{figure}

(2) The main tides generated by Jupiter and Saturn on the Sun are
characterized by two beating oscillations: the tidal oscillation associated
to the orbital period of Jupiter ($\sim$11.86 yr period) and the
spring tidal oscillation of Jupiter and Saturn ($\sim$9.93 yr period)
\citep{Brown,Scafetta2012c}. \citet{Scafetta2012b,Scafetta2012c}
used detailed spectral analysis of the sunspot monthly record since
1749 and showed that the $\sim$11 yr solar cycle is constrained by
the presence of two spectral peaks close to the two theoretical tidal
periods deduced from the orbits of Jupiter and Saturn: see Figure
6. These two frequencies modulate the main central cycle at $\sim$10.87
yr period. The beat generated by the superposition of the three harmonics
is characterized by four frequencies at about 61, 115, 130, and 983
yr periods which are typically observed in solar records \citep[e.g.: ][]{Ogurtsov,Scafetta2012b}.
\citet{Scafetta2012b} proposed a harmonic model for solar variability
based on three frequencies at periods of $\sim$9.93, $\sim$10.87
and $\sim$11.86 yr. The phases of the three harmonics were determined
from the conjunction date of Jupiter and Saturn (2000.475), the sunspot
record from 1749 to 2010 (2002.364) and the perihelion date of Jupiter
(1999.381), respectively. This simple three-frequency solar model
not only oscillates with a $\sim$11 yr cycle, as it should by mathematical
construction, it also manifests a complex multidecadal to millennia
beat modulation that has been shown to hindcast all major patterns
observed in both solar and climate records throughout the Holocene
\citep{Scafetta2012b}. For example, the model was shown to efficiently
hindcast: (1) the quasi millennia oscillation ($\sim$983 yr) found
in both climate and solar records \citep{Bond}; (2) the grand solar
minima during the last millennium such as the Oort, Wolf, Sp\"orer,
Maunder and Dalton minima; (3) seventeen $\sim$115 yr long oscillations
found in a detailed temperature reconstruction of the Northern Hemisphere
covering the last 2000 years; (4) the $\sim$59-63 yr oscillation
observed in the temperature record since 1850 and other features.
Scafetta's (2012b) three-frequency solar model forecasts that the
Sun will experience another moderate grand minimum during the following
decades and will return to a grand maximum in 2060's similar to the
grand maximum experienced in the 2000's: see Figure 7B.

\begin{figure}[!t]
\centering{}\includegraphics[width=0.8\textwidth]{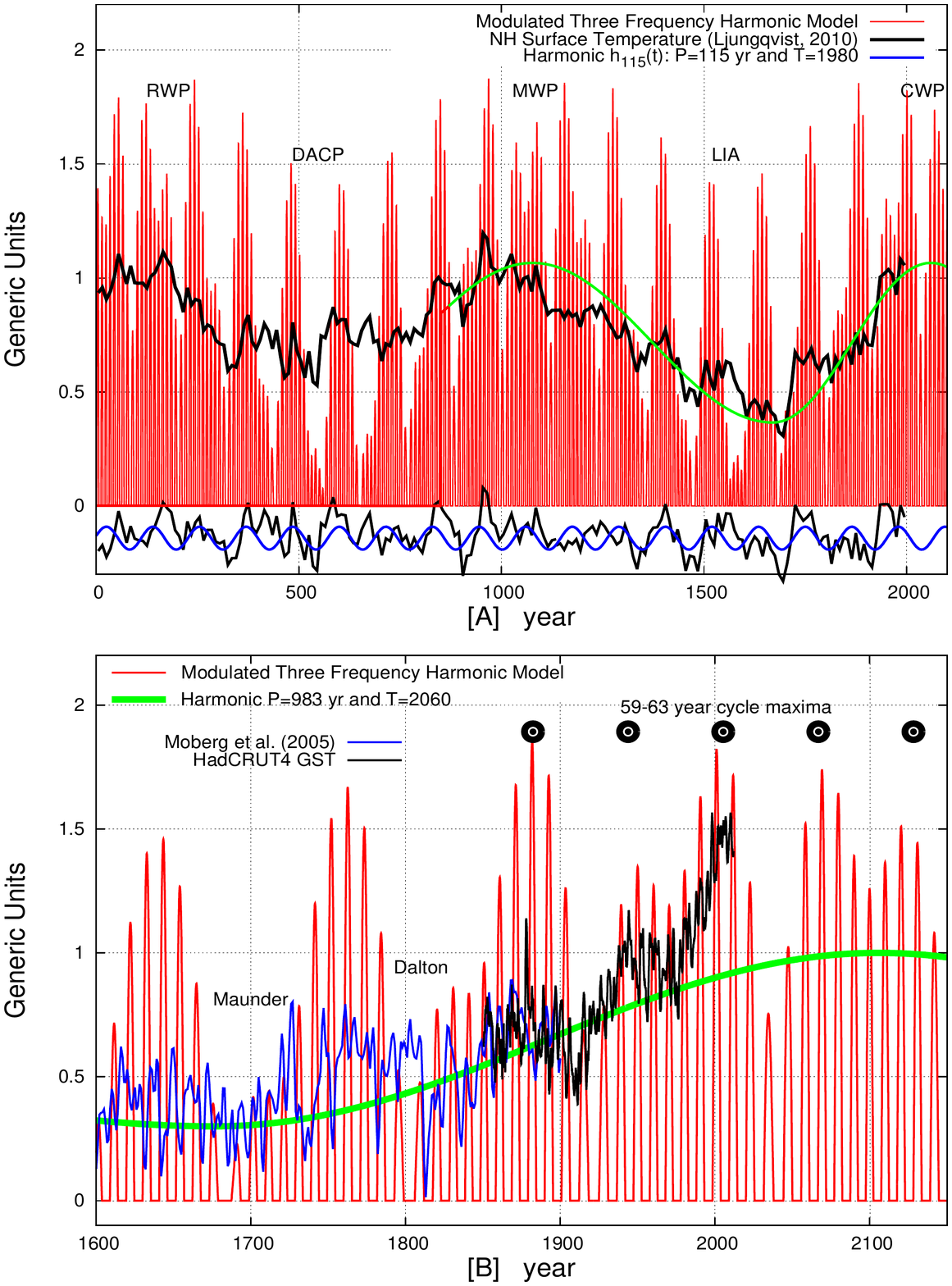}\caption{\citet{Scafetta2012b} three-frequency solar model (red). {[}A{]}
Against the Northern Hemisphere temperature reconstruction by \citet{Ljungqvist}
(black). The bottom depicts a filtering of the temperature reconstruction
(black) that highlights the 115-year oscillation (blue). {[}B{]} The
same solar model (red) is plotted against the HadCRUT4 global surface
temperature (black) merged in 1850\textendash{}1900 with the proxy
temperature model by \citet{Moberg} (blue). The green curves highlight
the quasi millennial oscillation with its skewness that approximately
reproduces the millennial temperature oscillation from 1700 to 2013.
Note the hindcast of the Maunder and Dalton solar minima and relative
cool periods, and the projected quasi 61-year oscillation from 1850
to 2150. Adapted from \citet{Scafetta2013a,Scafetta2013b}.}
\end{figure}

\citet{Solheim} observed that if the longer sunspot yearly resolved
record is used (1700-2012) the central spectral peak observed in Figure
6 at $\sim$10.87 yr could be split in two peaks as $\sim$11.01 yr
and $\sim$10.66 yr period. My own reanalysis of the periodogram of
the sunspot annual record since 1700 shows that the split produces
a secondary peak at $10.52\pm0.2$ yr and a main peak at $11.00\pm0.2$
yr. This result suggests that the central peak at $\sim$10.87 yr,
which was interpreted in \citet{Scafetta2012b,Scafetta2012c} as being
produced by an internal dynamo cycle, could indeed emerge from the
Venus-Earth-Jupiter recurrent cycles at $\sim$11.07 yr period plus
a possible minor cycle at $\sim$10.57 yr period. Figure 4 shows that
these two spectral peaks, plus another one at $\sim$11.26 yr period,
are among the planetary harmonics. This issue needs further analysis.
As for the ocean tidal system on Earth, it is possible that multiple
planetary oscillations regulate the $\sim$11 yr solar cycle.

The physical meaning of the three-frequency solar model is that solar
variability at the multidecadal to millennia scales is mostly determined
by the interference among the harmonic constituents that make up the
main $\sim$11 yr solar oscillation. When these harmonics interfere
destructively the Sun enters into a prolonged grand minimum; when
they interfere constructively the Sun experiences a grand maximum.
Additional oscillations at $\sim$45, $\sim$85, $\sim$170 and $\sim$210
yr period, also driven by the other two giant planets, Uranus and
Neptune (see Figure 4), have been observed in long solar and aurora
records \citep{Ogurtsov,Scafetta2012b,ScafettaWillson2013a}, but
not yet included to optimize the three-frequency solar model. 

\begin{figure}[!t]
\includegraphics[width=1\textwidth]{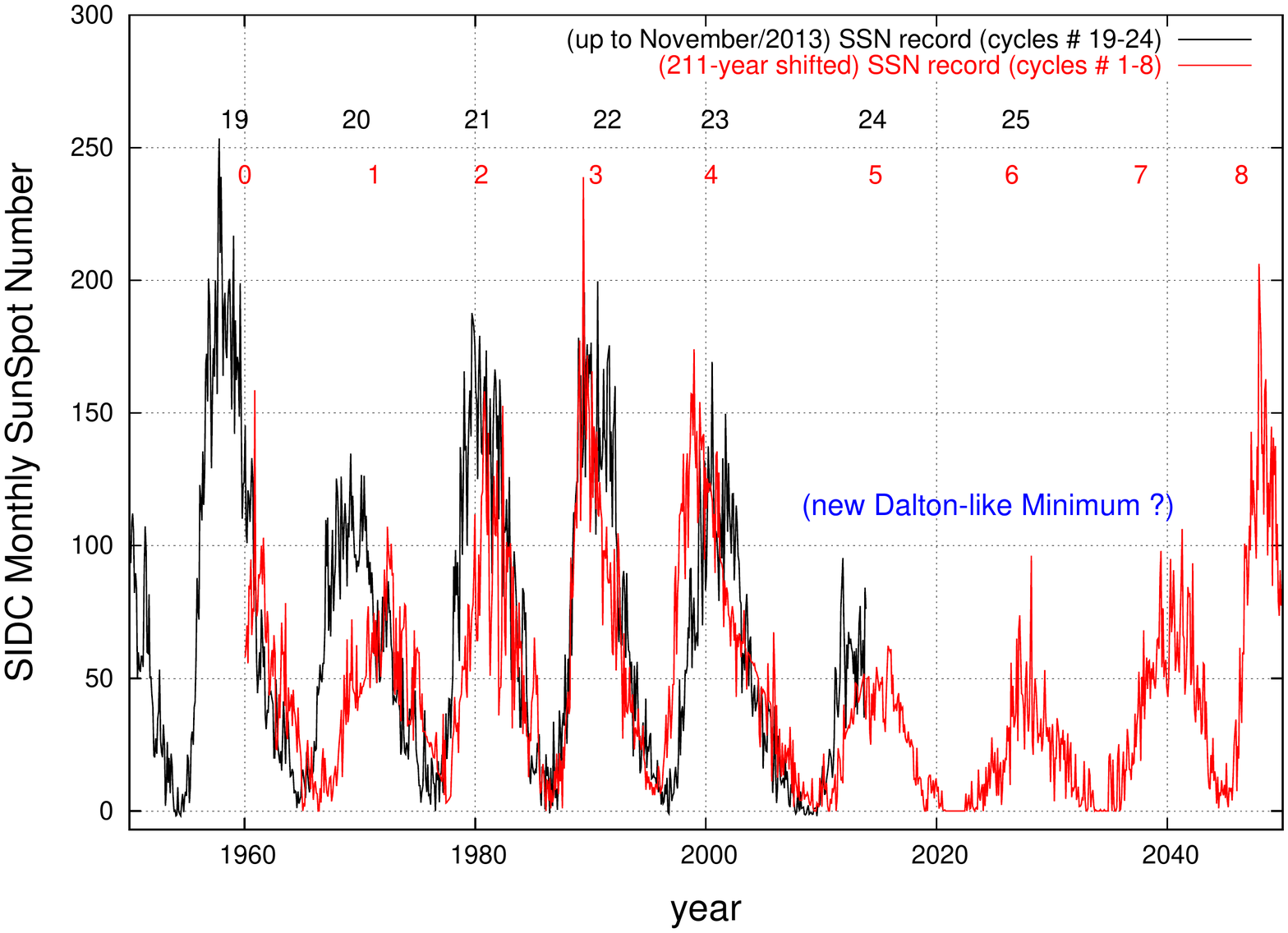}\caption{Comparison between latest sunspot cycles \#19-24 (black) and the sunspot
cycles \#1-5 (red) immediately preceding the Dalton Minimum (1790-1830).
A new Dalton-like solar minimum is likely approaching and may last
until 2045. The 211 yr temporal lag approximately corresponds to a
Suess - de Vries solar cycle, which approximately corresponds to the
\textasciitilde{}110 yr beat period between the $\sim$60 yr Jupiter-Saturn
beat (Figs. 2C and 4A) and the 84 yr Uranus orbital cycle. From \citet{Scafetta2012b}.}
\end{figure}

Note that the three-frequency solar model proposed by \citet{Scafetta2012b}
is a semi-empirical model because it is based on the two main physical
tidal harmonics generated by Jupiter and Saturn plus a statistically-estimated
central $\sim$11 yr solar harmonic. Therefore, this model is based
on both astronomical and empirical considerations, and its hindcasting
capability have been tested for centuries and millennia. Alternative
empirical models of solar variability directly based on long range
harmonics determined using power spectra and linear regressions of
solar records have been also proposed \citep[e.g.:][]{ScafettaWillson2013a,Solheim,Salvador,Steinhilber}.
However, models based on as many as possible astronomical and physical
considerations should be preferred to purely statistical or regressive
models because the former are characterized by a lower number of degrees
of freedom than the latter for the same number of modeled harmonics. 

The proposed semi-empirical and empirical harmonic solar models agree
about the fact that the Sun is entering into a period of grand minimum.
Indeed, the latest sunspot cycles \#19-24 are closely correlated to
the sunspot cycles \#1-5 immediately preceding the Dalton Minimum
(1790-1830): see Figure 8. \citet{Battistini} noted that the 11-year
solar cycle model proposed by \citet{Bendandi} based on the Venus-Earth-Jupiter
configuration is slightly out of phase with both the sunspot cycles
\#2-4 preceding the Dalton Minimum and with the sunspot cycles \#22-24.
This result may also be another evidence suggesting that the situation
preceding the Dalton Minimum is repeating today and could be anticipated
by a planetary configuration.

\section{Astronomically-based semi-empirical harmonic climate models}

As already understood since antiquity \citep[cf. ][]{Ptolemy}, \citet{Kepler1601}
recognized that the moon plays a crucial role in determining the ocean
tidal oscillations and, in doing so, he anticipated \citet{Newton}
in conceiving invisible forces (gravity and electromagnetism) that
could act at great distances. Kepler also argued that the climate
system could be partially synchronized to multiple planetary harmonics
\citep{Kepler1601,Kepler1606}. The main long scale harmonics that
Kepler identified were a $\sim$20 yr oscillation, a $\sim$60 yr
oscillation and a quasi millennial oscillation. These oscillations
were suggested by the conjunctions of Jupiter and Saturn and by historical
chronological considerations \citep{Kepler1606,Masar}: see Figure
2C. The quasi-millennial oscillation was associated to the slow rotation
of the \textit{trigon} of the conjunctions of Jupiter and Saturn,
and \citet{Kepler1606} claimed that this cycle was $\sim$800 yr
long. Kepler's calculations were based on the tropical orbital periods
of Jupiter and Saturn, which is how the orbits of Jupiter and Saturn
are seen from the Earth. However, using the sidereal orbital periods
this oscillation should be 850-1000 yr long \citep{Scafetta(2012a)},
as suggested in the power spectrum analysis shown in Figure 4. Since
antiquity equivalent climatic oscillations have been noted \citep{Iyengar,Masar,Temple}
and inserted in traditional calendars. For example, the Indian and
Chinese traditional calendars are based on a 60 yr cycle known in
the Indian tradition as the \textit{Brihaspati} (which means \textit{Jupiter})
cycle. 

The existence of climatic oscillations at about 10, 20, 60 and 1000
yr (and others) have been confirmed by numerous modern studies analyzing
various instrumental and proxy climatic records such as the global
surface temperature, the Atlantic Multidecadal Oscillation (AMO),
the Pacific Decadal Oscillation (PDO), the North Atlantic Oscillation
(NAO), ice core records, tree ring records, sea level records, fishery
records, etc. \citep[e.g.:][]{Bond,Chylek2011,Klyashtorin,Knudsen,Jevrejeva,Morner1989,Scafetta(2012a),Scafetta2013c,Wyatt}.
Indeed, numerous authors have also noted a correlation at multiple
scales between climate oscillations and planetary functions, for example
those related to the dynamics of the Sun relative to the barycenter
of the solar system \citep[e.g.:][]{Charvatova1997,Charvatova2013,Fairbridge,Jakubcova,Landscheidt,Scafetta2010,Scafetta2012b,Solheim}.

In particular, global surface temperature records, which are available
since 1850, present at least four major spectral peaks at periods
of about 9.1, 10-11, 20 and 60 yr, plus three minor peaks at about
12, 15 and 30 yr. See figure 1 in \citet{Scafetta2013b}, which is
partially reproduced in \citet{Solheim-b}. Sub-decadal astronomical
oscillations are also observed in climatic records \citet{Scafetta2010}.
In addition, multisecular and millennial oscillations (e.g. there
are major $\sim$115 yr and $\sim$983 yr oscillations and others)
can be deduced from paleoclimatic proxy temperature models. As also
shown in Figure 4, these oscillations can be associated with planetary
harmonics \citep{Scafetta2010,Scafetta2012b}. Astronomically-based
semi-empirical harmonic models to reconstruct and forecast climatic
changes are being proposed by several authors \citep[e.g.: ][]{Abdusamatov,Akasofu,Ludecke,Salvador,Scafetta2010,Scafetta(2012a),Scafetta2012b,Scafetta2012d,Scafetta2013a,Solheim}. 

For example, \citet{Scafetta2013b} proposed a semi-empirical harmonic
climate model based on astronomical oscillations plus an anthropogenic
and volcano contribution. In its latest form this model is made of
the following 6 astronomically deduced harmonics with periods of 9.1,
10.4, 20, 60, 115, 983 yr:

\begin{equation}
\begin{array}{lcl}
h_{9.1}(t) & = & 0.044\cos\left(2\pi\left(t-1997.82\right)/9.1\right)\\
h_{10.4}(t) & = & 0.030\cos\left(2\pi\left(t-2002.93\right)/10.4\right)\\
h_{20}(t) & = & 0.043\cos\left(2\pi\left(t-2001.43\right)/20\right)\\
h_{60}(t) & = & 0.111\cos\left(2\pi\left(t-2001.29\right)/60\right)\\
h_{115}(t) & = & 0.050\cos\left(2\pi\left(t-1980\right)/115\right)\\
h_{983}(t) & = & 0.350\cos\left(2\pi\left(t-2060\right)/760\right)
\end{array}\label{eq:mH}
\end{equation}
In the last equation a 760 yr period from 1680 to 2060 is used instead
of a 983 yr period because the millennial temperature oscillation
is skewed. While its maximum is predicted to occur in 2060, the minimum
occurs around 1680 during the Maunder Minimum (1645-1715). See Figure
7A above and Figure 8 in \citet{Humlum}.

The $9.1$ yr cycle was associated with a soli-lunar tidal oscillation
\citep[e.g.:][]{Scafetta2010,Scafetta2012d}. The rationale was that
the lunar nodes complete a revolution in 18.6 yr and the Saros soli\textendash{}lunar
eclipse cycle completes a revolution in 18 yr and 11 days. These two
cycles induce 9.3 yr and 9.015 yr tidal oscillations corresponding
respectively to the Sun\textendash{}Earth\textendash{}Moon and Sun\textendash{}Moon\textendash{}Earth
symmetric tidal configurations. Moreover, the lunar apsidal precession
completes one rotation in 8.85 yr causing a corresponding lunar tidal
cycle. The three cycles cluster between 8.85 yr and 9.3 yr periods
producing an average period around 9.06 yr. This soli\textendash{}lunar
tidal cycle has to peak in 1997\textendash{}1998 when the solar and
lunar eclipses occurred close to the equinoxes and the tidal torque
was stronger because acting on the equator. Indeed, the $\sim$9.1
yr temperature cycle was found to peak in 1997.82, as expected from
the soli-lunar cycle model \citep{Scafetta2012d}. 

\begin{figure}[!t]
\centering{}\includegraphics[width=1\textwidth]{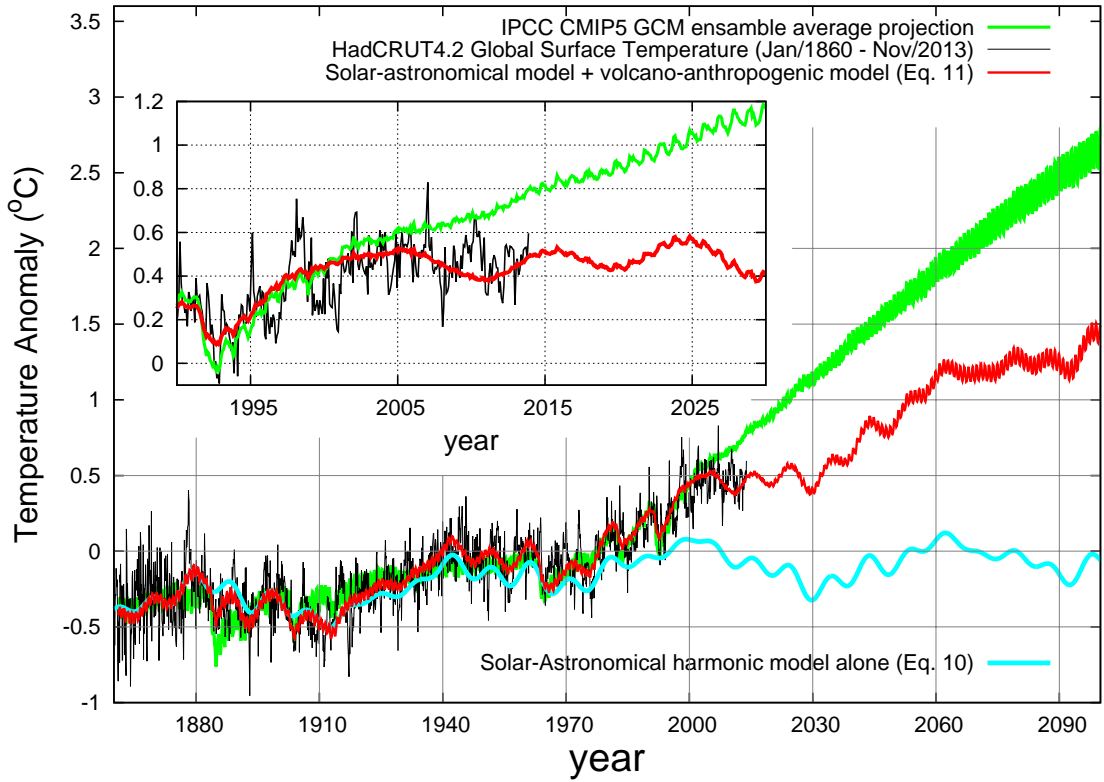}\caption{The semi-empirical model (Eq. \ref{eq:H(t)}) using $\beta=0.5$ (red)
attenuation of the CMIP5 GCM ensemble mean simulation vs. HadCRUT4
GST record from Jan/1860 to Nov/2013 (black). The cyan curve represents
the natural harmonic component alone (Eqs. \ref{eq:mH}). The green
curve represents the CMIP5 GCM average simulation used by the IPCC
in 2013. The model reconstructs the 20$^{th}$ century warming and
all decadal and multidecadal temperature patterns observed since 1860
significantly better than the GCM simulations such as the standstill
since $\sim$1997, which is highlighted in the insert. \citep[cf.][]{Scafetta2010,Scafetta2012d,Scafetta2013b}.}
\end{figure}

The other five oscillations of Eq. \ref{eq:mH} were deduced from
solar and planetary oscillations. The $10.4$ yr cycle appears to
be a combination of the $\sim$10 yr Jupiter-Saturn spring cycle and
the $\sim$11 yr solar cycle and peaks in 2002.93, that is $\sim$1
yr after the maximum of solar cycle 23, that occurred in $\sim2002$.
The $\sim$20 and $\sim$60 yr temperature cycles are synchronized
with the $\sim$20 and $\sim$60 yr oscillations of the speed of the
Sun relative to the center-of-mass of the solar system \citep{Scafetta2010}
and the $\sim$61 yr beat cycle of the Jupiter-Saturn tidal function,
which peaks around 1880s, 1940s and 2000s \citep{Scafetta2012b,Scafetta2012c}:
see also Figure 7B. I note, however, that \citet{Wilson2013b} proposed
a complementary explanation of the $\sim$60 yr climatic oscillation,
which would be caused by planetary induced solar activity oscillations
resonating with tidal oscillations associated to specific lunar orbital
variations synchronized with the motion of the Jovian planets. The
$\sim$115 and $\sim$983 yr oscillations are synchronized with both
the secular and millennial oscillations found in climatic and solar
proxy records during the Holocene \citep{Scafetta2012b}. The amplitude
of the millennial cycle is determined using modern paleoclimatic temperature
reconstructions \citep{Ljungqvist,Moberg}. The six oscillations of
Eq. \ref{eq:mH} are quite synchronous to the correspondent astronomical
oscillations: see Figure 7 and \citet{Scafetta2010,Scafetta2013b}.
Only the amplitudes of the oscillations are fully free parameters
that are determined by regression against the temperature record.
See \citet{Scafetta2010,Scafetta2012b,Scafetta2013b} for details. 

To complete the semi-empirical model, a contribution from anthropogenic
and volcano forcings was added. It could be estimated using the outputs
of typical general circulation models (GCMs) of the coupled model
intercomparison project 5 (CMIP5) simulations, $m(t),$ attenuated
by half, $\beta=0.5$ \citep{Scafetta2013b}. The attenuation was
required to compensate for the fact that the CMIP3 and CMIP5 GCMs
do not reproduce the observed natural climatic oscillations \citep[e.g.:][]{Scafetta2010,Scafetta2012d,Scafetta2013b}.
This operation was also justified on the ground that the CMIP5 GCMs
predict an almost negligible solar effect on climate change and their
simulations essentially model anthropogenic plus volcano radiative
effects alone. Finally, the adoption of $\beta=0.5$ was also justified
by the fact that numerous recent studies \citep[e.g.: ][]{Chylek2011,Chylek2008,Lewis,Lindzen and Choi,Ring,Scafetta2013b,Singer,Spencer,Zhou}
have suggested that the true climate sensitivity to radiative forcing
could be about half ($\sim$0.7-2.3 $^{o}C$ for $CO_{2}$ doubling)
of the current GCM estimated range ($\sim$1.5 to 4.5 $^{o}C$; \citet{IPCC}). 

Scafetta's (2013b) semi-empirical climate model is calculated using
the following formula:

\begin{equation}
H\left(t\right)=h_{9.1}(t)+h_{10.4}(t)+h_{20}(t)+h_{60}(t)+h_{115}(t)+h_{983}(t)+\beta*m(t)+const.\label{eq:H(t)}
\end{equation}
Figure 9 shows that the model (Eq. \ref{eq:H(t)}) successfully reproduces
all of the decadal and multi-decadal oscillating patterns observed
in the temperature record since 1850, including the upward trend and
the temperature standstill since 2000. On the contrary, the decadal
and multidecadal temperature oscillations and the temperature standstill
since $\sim$2000 are macroscopically missed by the CMIP5 GCM simulations
adopted by the Intergovernmental Panel on Climate Change \citep{IPCC}
\citep[cf. ][]{Scafetta2013b}. As Figure 9 shows, Eq. \ref{eq:H(t)}
projects a significant lower warming during the 21$^{st}$ century
than the CMIP5 average projection. 

Alternative empirical models for the global surface temperature have
been proposed by \citet{Scafetta2010,Scafetta(2012a),Scafetta2012d,Scafetta2013a},
\citet{Solheim-b}, \citet{Akasofu}, \citet{Abdusamatov}, \citet{Ludecke},
\citet{Vahrenholt} and others. These models are based on the common
assumption that the climate is characterized by specific quasi-harmonic
oscillations linked to astronomical-solar cycles. However, they differ
from each other in important mathematical details and physical assumptions.
These differences yield different performances and projections for
the 21$^{st}$ century. For example, Scafetta's (2010, 2012a, 2012d,
2013a, 2013b) models predict a temperature standstill until 2030s
and a moderate anthropogenic warming from 2000 to 2100 modulated by
natural oscillations such as the $\sim$60 yr cycle: see the red curve
in Figure 9. Scafetta's model takes into account that the natural
climatic variability, driven by a forecasted solar minimum similar
to a moderate Dalton solar minimum or to the solar minimum observed
during $\sim$1910 (see Figures 7B and 8) would yield a global cooling
of $\sim$0.4 $^{o}C$ from $\sim$2000 to $\sim$2030 (see cyan curve
in Figure 9), but this natural cooling would be mostly compensated
by anthropogenic warming as projected throughout the 21$^{st}$ century
by Scafetta's $\beta$-attenuated model (see Eq. \ref{eq:H(t)}).
Although with some differences, the climatic predictions of \citet{Solheim-b},
\citet{Akasofu} and \citet{Vahrenholt} look quite similar: they
predict a steady to moderate global cooling from 2000 to 2030 and
a moderate warming for 2100 modulated by a $\sim$60 yr cycle. On
the contrary, \citet[figure 8]{Abdusamatov} predicted an imminent
cooling of the global temperature beginning since the year 2014 that
will continue throughout the first half of the 21$^{st}$ century
that would yield a Little-Ice-Age period from $\sim$2050 to $\sim$2110
when the temperature would be $\sim$1.2 $^{o}C$ cooler than the
2000-2010 global temperature. Abdusamatov's predicted strong cooling
would be induced by an approaching Maunder-like solar minimum period
that would occur during the second half of the 21$^{st}$ century.
\citet{Steinhilber} also predicted a grand solar minimum occurring
during the second half of the 21$^{st}$ century, but it would be
quite moderate and more similar to the solar minimum observed during
$\sim$1910; thus, this solar minimum will not be as deep as the Maunder
solar minimum of the 17$^{th}$ century.

An analysis and comparison of the scientific merits of each proposed
harmonic constituent solar and climate model based on astronomical
oscillations elude the purpose of this paper and it is left to the
study of the reader. In general, harmonic models based only on statistical,
Fourier and regression analysis may be misleading if the harmonics
are not physically or astronomically justified. Nonetheless, harmonic
constituent models can work exceptionally well in reconstructing and
forecasting the natural variability of a system if the dynamics of
the system is sufficiently harmonic and the constituent physical/astronomical
harmonics are identified with great precision. For example, the astronomically-based
harmonic constituent models currently used to predict the ocean tides
are the most accurate predictive geophysical models currently available
(\citet{Doodson}; \href{http://en.wikipedia.org/wiki/Theory_of_tides}{http://en.wikipedia.org/wiki/Theory\_{}of\_{}tides}). 

\citet{Scafetta2012b,Scafetta2012d,Scafetta2013b} carefully tested
his solar and climate models based on astronomical oscillations using
several hindcasting procedures. For example, the harmonic solar model
was tested in its ability to hindcast the major solar patterns during
the Holocene and the harmonic climate model was calibrated during
the period 1850-1950 and its performance to obtain the correct 1950-2010
patterns was properly tested, and vice versa. Future observations
will help to better identify and further develop the most reliable
harmonic constituent climate model based on astronomical oscillations.

\section{Conclusion}

Pythagoras of Samos \citep{Pliny} proposed that the Sun, the Moon
and the planets all emit their own unique \textit{hum} \citep[orbital resonance; cf. ][]{Tattersall}
based on their orbital revolution, and that the quality of life on
Earth reflects somehow the tenor of the celestial \textit{sounds}
(from \href{http://en.wikipedia.org/wiki/Musica_universalis}{http://en.wikipedia.org}).
This ancient philosophical concept is known as \textit{musica universalis}
(universal music or music of the spheres). However, it is with Copernicus'
heliocentric revolution that the harmonic structure of the solar system
became clearer. \citet{Kepler1596,Kepler1619} strongly advocated
the \textit{harmonices mundi} (the harmony of the world) concept from
a scientific point of view. 

Since the $17^{th}$ century scientists have tried to disclose the
fundamental mathematical relationships that describe the solar system.
Interesting resonances linking the planets together have been found.
I have briefly discussed the Titius-Bode rule and other resonant relationships
that have been proposed during the last centuries. In addition, planetary
harmonics have been recently found in solar and climate records, and
semi-empirical models to interpret and reconstruct the climatic oscillations,
which are not modeled by current GCMs, have been proposed \citep[e.g.][]{Scafetta2013b}. 

How planetary harmonics could modulate the Sun and the climate on
the Earth is still unknown. Some papers have noted that tidal torquing
function acting upon hypothesized distortions in the Sun\textquoteright{}s
tachocline present planetary frequencies similar to those found in
solar proxy and climate records \citep[e.g.,][]{Abreu,Wilson}. However,
whether planetary gravitational forces are energetically sufficiently
strong to modulate the Sun's activity in a measurable way remains
a serious physical problem and reason of skepticism. Also basic Newtonian
physics such a simple evaluations of tidal accelerations on just the
Sun's tachocline does not seem to support the theory due to the fact
that planetary tidal accelerations in the Sun seem too small (just
noise) compared to the strengths of the typical convective accelerations
\citep{Callebaut}. 

However, the small gravitational perturbation that the Sun is experiencing
are harmonic, and the Sun is a powerful generator of energy very sensitive
to gravitational and electromagnetic variations. Thus, the Sun's internal
dynamics could synchronize to the frequency of the external forcings
and could work as a huge amplifier and resonator of the tenuous gravitational
\textit{music} generated by the periodic synchronized motion of the
planets. \citet{Scafetta2012c} proposed a physical amplification
mechanism based on the mass-luminosity relation. In Scafetta's model
the Sun's tachocline would be forced mostly by an oscillating luminosity
signal emerging from the solar interior \citep[cf.:][]{Wolff}. The
amplitude of the luminosity anomaly signal driven by the planetary
tides, generated in the Sun's core and quickly propagating as acoustic-like
waves in the radiative zone into the Sun's tachocline has to oscillate
with the tidal and torquing planetary gravitational frequencies because
function of the gravitational tidal potential energy dissipated in
the solar interior. The energetic strength of this signal was estimated
and found to be sufficiently strong to synchronize the dynamics of
the Sun's tachocline and, consequently, of the Sun's convective zone.
The quasi harmonic and resonant structure observed in the solar system
should further favor the emergence of collective synchronization patterns
throughout the solar system and activate amplification mechanisms
in the Sun and, consequently, in the Earth's climate. 

Although a comprehensive physical explanation has not been fully found
yet, uninterrupted aurora records, solar records, and long solar proxy
records appear to be characterized by astronomical harmonics from
the monthly to the millennial time scales, and the same harmonics
are also present in climate records, as found by numerous authors
since the $19^{th}$ century \citep[e.g.: ][]{Wolf1859,Brown,Abreu,Charvatova,Charvatova2013,Fairbridge,Hung,Jakubcova,Jose,Scafetta2010,Scafetta(2012a),Scafetta2012b,Scafetta2012c,Salvador,Scafetta2012d,Scafetta2013b,ScafettaWillson2013b,ScafettaWillson2013a,ScafettaWillson2013c,Sharp,Solheim,Tan,Wilson2011,Wilson,Wolff}.
Thus, gravitational and electromagnetic planetary forces should modulate
both solar activity and, directly or indirectly, the electromagnetic
properties of the heliosphere. The climate could respond both to solar
luminosity oscillations and to the electromagnetic oscillations of
the heliosphere, and synchronize to them. The electromagnetic oscillations
of the heliosphere and the interplanetary electric field could directly
influence the Earth's cloud system through a modulation of cosmic
ray and solar wind causing oscillations in the terrestrial albedo,
which could be sufficiently large (about 1-3\%) to cause the observed
climatic oscillations \citep[e.g: ][]{Morner2013,Scafetta(2012a),Scafetta2013b,Svensmark,Tinsley,Voiculescu}. 

Although the proposed rules and equations are not \textit{perfect}
yet, the results do support the idea that the solar system is highly
organized in some form of complex resonant and synchronized structure.
However, this state is dynamical and is continuously perturbed by
chaotic variability, as it should be physically expected. Future research
should investigate planets-Sun and space-climate coupling mechanisms
in order to develop more advanced and efficient analytical and semi-empirical
solar and climate models. Perhaps, a harmonic set made of the planetary
harmonics listed in Figure 4 plus the beat harmonics generated by
the solar synchronization \citep[e.g. ][]{Scafetta2012b} plus the
harmonics deducible from the soli-lunar tides \citep[e.g. ][]{Wang2012}
constitutes the harmonic constituent group that is required for developing
advanced astronomically-based semi-empirical harmonic climate models.

As Pythagoras, Ptolemy, Kepler and many civilizations conjectured
since antiquity, solar and climate forecasts and projections based
on astronomical oscillations appear physically possible. Advancing
this scientific research could greatly benefit humanity.

\section*{Acknowledgment}

The author thanks R. C. Willson (ACRIM science team) for support,
the referees for useful suggestions and the editors for having organized
the Special Issue on \textit{``Pattern in solar variability, their
planetary origin and terrestrial impacts}'' \citep{Morneretal2013}
where 10 authors try to further develop the ideas about the planetary-solar-terrestrial
interaction.


\begin{thebibliography}{Charv\'atov\'a and Hejda(2013)}
\bibitem[Abdusamatov(2013)]{Abdusamatov}Abdusamatov, Kh. I.: Grand
minimum of the total solar irradiance leads to the little ice age.
Geology \& Geosciences 28(2), 62-68, 2013.

\bibitem[Abreu et al.(2012)]{Abreu} Abreu, J. A., Beer, J., Ferriz-Mas,
A., McCracken, K. G., Steinhilber, F.: Is there a planetary influence
on solar activity? Astronony \& Astrophysics 548, A88, doi: 10.1051/0004-6361/201219997,
2012.

\bibitem[Akasofu(2013)]{Akasofu}Akasofu, S.-I.: On the Present Halting
of Global Warming. Climate. 1(1), 4-11, 2013.

\bibitem[Battistini(2011)]{Battistini}Battistini, A.: Il ciclo undecennale
del sole secondo Bendandi (The 11-year solar cycle according to Bendandi).
New Ice Age, \href{http://daltonsminima.altervista.org/?p=8669}{http://daltonsminima.altervista.org/?p=8669},
2011.

\bibitem[Bendandi(1931)]{Bendandi} Bendandi, R.: Un principio fondamentale
dell\textquoteright{}Universo (A fundamental principle of the Universe).
(Faenza, Osservatorio Bendandi), 1931.

\bibitem[Bigg(1967)]{Bigg}Bigg, E.K.: Influence of the planet Mercury
on sunspots. Astronomical Journal 72, 463\textendash{}466, 1967.

\bibitem[Brown(1900)]{Brown} Brown, E. W.: A Possible Explanation
of the Sun-spot Period. Monthly Notices of the Royal Astronomical
Society 60, 599-606, 1900.

\bibitem[Bode(1772)]{Bode} Bode, J. E.: Anleitung zur Kenntnis des
gestirten Himmels, 2nd edn (Hamburg), p. 462, 1772.

\bibitem[Bollinger(1952)]{Bollinger}Bollinger, C. J.: A 44.77 year
Jupiter-Earth-Venus configuration Sun-tide period in solar-climate
cycles. Academy of Science for 1952 \textendash{} Proceedings of the
Oklahoma, 307\textendash{}311, 1952. (\href{http://digital.library.okstate.edu/oas/oas_pdf/v33/v307_311.pdf}{http://digital.library.okstate.edu/oas/oas\_{}pdf/v33/v307\_{}311.pdf})

\bibitem[Bond et al.(2001)]{Bond}Bond, G., Kromer, B., Beer, J.,
Muscheler, R., Evans, M.N., Showers,W., Hoffmann, S., Lotti- Bond,
R., Hajdas, I., Bonani, G., 2001. Persistent solar influence on North
Atlantic climate during the Holocene. Science 294, 2130\textendash{}2136,
2001.

\bibitem[Bucha et al.(1985)]{Bucha} Bucha, V., Jakubcov\'a, I.,
Pick, M., 1985. Resonance frequencies in the Sun\textquoteright{}s
motion. Studia Geophysica et Geodaetica 29, 107\textendash{}111, 1985.

\bibitem[Callebaut et al.(2012)]{Callebaut}Callebaut, D. K., de Jager,
C., and Duhau, S.: The influence of planetary attractions on the solar
tachocline, J. Atmos. Sol.-Terr. Phy., 80, 73\textendash{}78, 2012.

\bibitem[Charv\'atov\'a(1997)]{Charvatova1997}Charv\'atov\'a, I.:
Solar-terrestrial and climatic phenomena in relation to solar inertial
motion. Surveys in Geophys., 18, 131-146, 1997.

\bibitem[Charv\'atov\'a(2009)]{Charvatova} Charv\'atov\'a, I.:
Long-term predictive assessments of solar and geomagnetic activities
made on the basis of the close similarity between the solar inertial
motions in the intervals 1840\textendash{}1905 and 1980\textendash{}2045.
New Astronomy 14, 25-30, 2009.

\bibitem[Charv\'atov\'a and Hejda(2013)]{Charvatova2013}Charv\'atov\'a,
I. and Hejda, P.: Responses of the basic cycle of 178.7 years and
2402 years in solar-terrestrial phenomena during Holocene, Pattern
Recogn. Phys., 1, in press.

\bibitem[Chylek and Lohmann(2008)]{Chylek2008} Chylek, P., Lohmann,
U.: Aerosol radiative forcing and climate sensitivity deduced from
the Last Glacial Maximum to Holocene transition. Geophys. Res. Lett.
35 (4), L04804, 2008.

\bibitem[Chylek et al.(2011)]{Chylek2011}Chylek, P., Folland, C.K.,
Dijkstra, H.A., Lesins, G., Dubey,M.K.: Ice-core data evidence for
a prominent near 20 year time-scale of the Atlantic Multidecadal Oscillation.
Geophys. Res. Lett. 38, L13704, 2011.

\bibitem[Copernicus(1543)]{Copernicus} Copernicus, N.: De revolutionibus
orbium coelestium. (Johannes Petreius), 1543.

\bibitem[de Jager and Versteegh(2005)]{deJager} de Jager, C. and
Versteegh, J. M.: Do planetary motions drive solar variability?, Sol.
Phys., 229, 175\textendash{}179, 2005.

\bibitem[Doodson(1921)]{Doodson} Doodson, A. T.: The harmonic development
of the tide-generating potential, Proceedings of the Royal Society
of London. Series A 100 (704), 305-329, 1921.

\bibitem[Dreyer(1912)]{Dreyer} Dreyer, J. L. E.: ``The Scientific
Papers of Sir William Herschel'', ed. by J.L.E. Dreyer (2 vols, London),
vol. 1, p. xxviii. 1912.

\bibitem[Dubrulle and Graner(1994a)]{Dubrullea}Dubrulle, B., Graner,
F.: Titius-Bode laws in the solar system. Part I: Scale invariance
explains everything. Astronomy and Astrophysics 282, 262\textendash{}268,
1994a.

\bibitem[Dubrulle and Graner(1994b)]{Dubrulleb}Dubrulle, B., Graner,
F.: Titius\textendash{}Bode laws in the solar system. Part II: Build
your own law from disk models. Astronomy and Astrophysics 282, 269\textendash{}276,
1994b.

\bibitem[Ebner(2011)]{Ebner}Ebner, J. E.: Gravity, Rosettes and Inertia.
The General Science Journal, October 16, pp. 1-11, 2011. \href{http://gsjournal.net/Science-Journals/Essays/View/3700}{http://gsjournal.net/Science-Journals/Essays/View/3700}

\bibitem[Fairbridge and Shirley(1987)]{Fairbridge} Fairbridge, R.
W., Shirley, J. H.: Prolonged minima and the 179-year cycle of the
solar inertial motion. Solar Physics 10, 191-210, 1987.

\bibitem[Geddes and King-Hele(1983)]{Geddes} Geddes, A. B. and King-Hele,
D. G.: Equations for Mirror Symmetries Among the Distances of the
Planets, Quarterly Journal of the Royal Astronomical Society, 24,
10-13, 1983.

\bibitem[Goldreich and Peale(1966a)]{Goldreich} Goldreich P., Peale,
S. J.: Resonant Rotation for Venus? Nature 209, 1117-1118, 1966a. 

\bibitem[Goldreich and Peale(1966b)]{Goldreichb} Goldreich P., Peale,
S. J.: Resonant Spin States in the Solar System. Nature 209, 1178-1179,
1966b. 

\bibitem[Humlum et al.(2011)]{Humlum}Humlum, O., Solheim, J.-E.,
Stordahl, K.: Identifying natural contributions to late Holocene climate
change. Glob. Planet. Chang. 79, 145\textendash{}156, 2011.

\bibitem[Hung(2007)]{Hung} Hung, C.-C.: Apparent Relations Between
Solar Activity and Solar Tides Caused by the Planets, NASA report/TM-
2007-214817, available at: \href{http://ntrs.nasa.gov/search.jsp?R=20070025111}{http://ntrs.nasa.gov/search.jsp?R=20070025111},
2007.

\bibitem[IPCC(2013)]{IPCC} Intergovernmental Panel on Climate Change
(IPCC) Fifth Assessment Report (AR5), \textit{Climate Change 2013:
The Physical Science Basis}. 2013.

\bibitem[Iyengar(2009)]{Iyengar} Iyengar, R. N.: Monsoon rainfall
cycles as depicted in ancient Sanskrit texts. Curr. Sci. 97, 444\textendash{}447,
2009.

\bibitem[Jakubcov\'a and Pick(1986)]{Jakubcova}Jakubcov\'a, I.,
Pick, M.: The planetary system and solar-terrestrial phenomena. Studia
Geophysica et Geodaetica 30, 224\textendash{}235, 1986.

\bibitem[Jelbring(2013)]{Jelbring2013a} Jelbring, H.: Celestial commensurabilities:
some special cases, Pattern Recogn. Phys., 1, 143-146, doi:10.5194/prp-1-143-2013,
2013.

\bibitem[Jevrejeva et al.(2008)]{Jevrejeva}Jevrejeva, S., Moore,
J.C., Grinsted, A., Woodworth, P.: Recent global sea level acceleration
started over 200 years ago? Geophys. Res. Lett. 35, L08715, 2008.

\bibitem[Jiang et al.(2007)]{Jiang}Jiang, J., Chatterjee, P., Choudhuri,
A.R.: Solar activity forecast with a dynamo model. Monthly Notices
of the Royal Astronomical Society 381, 1527\textendash{}1542, 2007.

\bibitem[Jose(1965)]{Jose}Jose, P.D.: Sun\textquoteright{}s motion
and sunspots. Astronomical Journal 70, 193\textendash{}200, 1965.

\bibitem[Kepler(1596)]{Kepler1596}Kepler, J.: Mysterium Cosmographicum
(The Cosmographic Mystery). 1596.

\bibitem[Kepler(1601)]{Kepler1601} Kepler, J.: On the more certain
fundamentals of astrology. 1601. In: Brackenridge, J.B., Rossi, M.A.
(Eds.), Proceedings of the American Philosophical Society, 123(2),
pp. 85\textendash{}116 (1979).

\bibitem[Kepler(1606)]{Kepler1606}Kepler, J.: De Stella Nova in Pede
Serpentarii (\textquotedbl{}On the new star in Ophiuchus's foot\textquotedbl{}).
1606.

\bibitem[Kepler(1609)]{Kepler1609} Kepler, J.: Astronomia nova (New
Astronomy). 1609. Translated by Donahue W.H., (Cambridge: Cambridge
Univ. Pr.) 1992.

\bibitem[Kepler(1619)]{Kepler1619}Kepler, J.: Harmonices mundi (The
Harmony of the World). 1619. Translated by Field J.. (The American
Philosophical Society) 1997.

\bibitem[Klyashtorin et al.(2009)]{Klyashtorin}Klyashtorin, L.B.,
Borisov, V., Lyubushin, A.: Cyclic changes of climate and major commercial
stocks of the Barents Sea. Mar. Biol. Res. 5, 4\textendash{}17, 2009.

\bibitem[Knudsen(2011)]{Knudsen}Knudsen, M.F., Seidenkrantz, M.-S.,
Jacobsen, B.H., Kuijpers, A.: Tracking the Atlantic Multidecadal Oscillation
through the last 8,000 years. Nat. Commun. 2, 178, 2011.

\bibitem[Landscheidt(1989)]{Landscheidt} Landscheidt, T.: Sun-Earth-Man,
a mesh of cosmic oscillations. (Urania Trust) 1989.

\bibitem[Lewis(2013)]{Lewis}Lewis, N.: An objective Bayesian, improved
approach for applying optimal fingerprint techniques to estimate climate
sensitivity. J. Clim. http://dx.doi.org/10.1175/ JCLI-D-12-00473.1,
2013.

\bibitem[Lindzen and Choi(2011)]{Lindzen and Choi} Lindzen, R.S.,
Choi, Y.-S.: On the observational determination of climate sensitivity
and its implications. Asia Pac. J. Atmos. Sci. 47, 377\textendash{}390,
2011.

\bibitem[Ljungqvist(2010)]{Ljungqvist}Ljungqvist, F.C.: A new reconstruction
of temperature variability in the extratropical Northern Hemisphere
during the last two millennia. Geogra-fiska Annaler Series A, 92,
pp. 339\textendash{}351, 2010.

\bibitem[L\"udecke et al.(2013)]{Ludecke} L\"udecke, H.-J., Hempelmann,
A., Weiss, C. O.: Multi-periodic climate dynamics: spectral analysis
of long-term instrumental and proxy temperature records. Clim. Past,
9, 447\textendash{}452, 2013.

\bibitem[Ma'Sar(IX century)]{Masar} Ma'Sar, A.: On Historical Astrology\textemdash{}The
Book of Religions and Dynasties (On the Great Conjunctions), $9^{th}$
century. Translated by Yamamoto, K., Burnett, C. (Brill Academic Publishers)
2000.

\bibitem[Moberg et al.(2005)]{Moberg}Moberg, A., Sonechkin, D.M.,
Holmgren, K., Datsenko, N.M., Karlén, W.: Highly variable Northern
Hemisphere temperatures reconstructed from low- and highresolution
proxy data. Nature 433, 613\textendash{}617, 2005.

\bibitem[Molchanov(1968)]{Molchanov1968} Molchanov, A. M.: The Resonant
Structure of the Solar System: The Law of Planetary Distances. Icarus
8, 203-215, 1968.

\bibitem[Molchanov(1969a)]{Molchanov1969a} Molchanov, A. M.: The
Reality of Resonances in the Solar System. Icarus ll, 104-110, 1969a.

\bibitem[Molchanov(1969b)]{Molchanov1969b} Molchanov, A. M.: Resonances
in Complex Systems: A Reply to Critiques. Icarus 11, 95-103, 1969b.

\bibitem[M\"orner(1989)]{Morner1989} M\"orner, N.-A.: Changes in
the Earth's rate of rotation on an El Nino to century basis. In: Lowes,
F.J., et al. (Ed.), Geomagnetism and Paleomagnetism. Kluwer Acad.
Publ., pp. 45\textendash{}53, 1989.

\bibitem[M\"orner(2013)]{Morner2013}M\"orner, N.-A.: Planetary beat
and solar\textendash{}terrestrial responses, Pattern Recogn. Phys.,
1, 107-116, doi:10.5194/prp-1-107-2013, 2013.

\bibitem[M\"orner et al.(2013)]{Morneretal2013} M\"orner, N.-A.,
Tattersall, R., and Solheim, J.-E.: Preface: Pattern in solar variability,
their planetary origin and terrestrial impacts, Pattern Recogn. Phys.,
1, 203-204, doi:10.5194/prp-1-203-2013, 2013.

\bibitem[Newton(1687)]{Newton} Newton, I.: Philosophiæ Naturalis
Principia Mathematica (Mathematical Principles of Natural Philosophy).
1687. 

\bibitem[Ogurtsov et al.(2002)]{Ogurtsov} Ogurtsov, M. G., Nagovitsyn,
Y. A., Kocharov, G. E., Jungner, H.: Long-period cycles of the sun's
activity recorded in direct solar data and proxies. Sol. Phys. 211,
371\textendash{}394, 2002.

\bibitem[Piazzi(1801)]{Piazzi} Piazzi, G.: Risultati delle Osservazioni
della Nuova Stella (Palermo) pp. 3-6, 1801. 

\bibitem[Pikovsky(2001)]{Pikovsky} Pikovsky, A., Rosenblum, M., Kurths,
J.: Synchronization, a universal concept in nonlinear science. (Cambridge
University Press, Cambridge UK) 2001.

\bibitem[Pliny the Elder(77AD)]{Pliny}Pliny the Elder: Natural History,
books I-II, 77AD. Translated by Rackham, H. (Harvard University Press,
1938).

\bibitem[Press et al.(1997)]{Press}Press, W. H., Teukolsky, S. A.,
Vetterling, W. T., and Flannery, B. P.: Numerical Recipes in C, 2nd
Edn., Cambridge University Press, 1997.

\bibitem[Ptolemy(II century)]{Ptolemy} Ptolemy, C.: Tetrabiblos:
on the influence of the stars, 2$^{nd}$ century. Translated by Ashmand,
J. M. (Astrology Classics, Bel Air, MD), 1980.

\bibitem[Ring et al.(2012)]{Ring} Ring, M.J., Lindner, D., Cross,
E.F., Schlesinger, M.E., 2012. Causes of the global warming observed
since the 19th century. Atmos. Clim. Sci. 2 (4), 401\textendash{}415.

\bibitem[Salvador(2013)]{Salvador}Salvador, R. J.: A mathematical
model of the sunspot cycle for the past 1000 yr, Pattern Recogn. Phys.,
1, 117-122, doi:10.5194/prp-1-117-2013, 2013.

\bibitem[Scafetta(2010)]{Scafetta2010} Scafetta, N.:. Empirical evidence
for a celestial origin of the climate oscillations and its implications.
J. Atmos. Sol.-Terr. Phy., 72, 951\textendash{}970, doi:10.1016/j.jastp.2010.04.015,
2010.

\bibitem[Scafetta(2012a)]{Scafetta(2012a)} Scafetta, N.: A shared
frequency set between the historical mid-latitude aurora records and
the global surface temperature, J. Atmos. Sol.-Terr. Phy., 74, 145\textendash{}163,
doi:10.1016/j.jastp.2011.10.013, 2012a.

\bibitem[Scafetta(2012b)]{Scafetta2012b} Scafetta, N.: Multi-scale
harmonic model for solar and climate cyclical variation throughout
the Holocene based on Jupiter-Saturn tidal frequencies plus the 11
yr solar dynamo cycle, J. Atmos. Sol.-Terr. Phy., 80, 296\textendash{}311,
doi:10.1016/j.jastp.2012.02.016, 2012b. 

\bibitem[Scafetta(2012c)]{Scafetta2012c}Scafetta, N.: Does the Sun
work as a nuclear fusion amplifier of planetary tidal forcing? A proposal
for a physical mechanism based on the mass-luminosity relation, J.
Atmos. Sol.-Terr. Phy., 81\textendash{}82, 27\textendash{}40, doi:10.1016/j.jastp.2012.04.002,
2012c.

\bibitem[Scafetta(2012d)]{Scafetta2012d}Scafetta N.: Testing an astronomically
based decadal-scale empirical harmonic climate model versus the IPCC
(2007) general circulation climate models. J. Atmos. Sol.-Terr. Phy.,
80, 124-137, doi:10.1016/j.jastp.2011.12.005, 2012d.

\bibitem[Scafetta(2013a)]{Scafetta2013a}Scafetta N.: Solar and planetary
oscillation control on climate change: hind-cast, forecast and a comparison
with the CMIP5 GCMs. Energy \& Environment 24(3-4), 455\textendash{}496,
doi:10.1260/0958-305X.24.3-4.455, 2013a.

\bibitem[Scafetta(2013b)]{Scafetta2013b}Scafetta, N.: Discussion
on climate oscillations: CMIP5 general circulation models versus a
semi-empirical harmonic model based on astronomical cycles. Earth-Science
Reviews 126, 321-357, doi:10.1016/j.earscirev.2013.08.008, 2013b.

\bibitem[Scafetta(2013c)]{Scafetta2013c}Scafetta, N.: Multi-scale
dynamical analysis (MSDA) of sea level records versus PDO, AMO, and
NAO indexes. Climate Dynamics. http://dx.doi.org/10.1007/s00382-013-1771-3,
2013c. (in press).

\bibitem[Scafetta and Willson(2013a)]{ScafettaWillson2013a} Scafetta,
N. and Willson, R. C.: Planetary harmonics in the historical Hungarian
aurora record (1523\textendash{}1960), Planet. Space Sci., 78, 38\textendash{}44,
doi:10.1016/j.pss.2013.01.005, 2013a. 

\bibitem[Scafetta and Willson(2013b)]{ScafettaWillson2013b} Scafetta,
N. and Willson, R. C.: Empirical evidences for a planetary modulation
of total solar irradiance and the TSI signature of the 1.09 yr Earth-Jupiter
conjunction cycle, Astrophys. Space Sci., 348, 25-39 doi:10.1007/s10509-013-1558-3,
2013b.

\bibitem[Scafetta and Willson(2013c)]{ScafettaWillson2013c} Scafetta,
N. and Willson, R. C.: Multiscale comparative spectral analysis of
satellite total solar irradiance measurements from 2003 to 2013 reveals
a planetary modulation of solar activity and its nonlinear dependence
on the 11 yr solar cycle, Pattern Recogn. Phys., 1, 123-133, doi:10.5194/prp-1-123-2013,
2013c.

\bibitem[Scharf(2010)]{Scharf} Scharf, C.A.: Possible Constraints
on Exoplanet Magnetic Field Strengths from Planet\textendash{}star
Interaction. The Astrophysical Journal 722, 1547\textendash{}1555,
2010.

\bibitem[Sharp(2013)]{Sharp} Sharp, G.J.: Are Uranus \& Neptune Responsible
for Solar Grand Minima and Solar Cycle Modulation? Int. J. Astron.
Astrophys. 3, 260-273, 2013.

\bibitem[Shirley(1990)]{Shirley}Shirley, J. H., Sperber, K. R., Fairbridge,
R. W.: Sun's internal motion and luminosity. Solar Physics 127, 379-392,
1990.

\bibitem[Shkolnik et al.(2003)]{Shkolnik2003}Shkolnik, E., Walker,
G. A. H., Bohlender D. A.: Evidence for Planet-induced Chromospheric
Activity on HD 179949. ApJ, 597, 1092-1096, 2003.

\bibitem[Shkolnik et al.(2005)]{Shkolnik2005}Shkolnik, E., Walker,
G. A. H., Bohlender, D. A., Gu, P.-G., and Kurster, M.: Hot Jupiters
and Hot Spots: The Short- and Long-Term Chromospheric Activity on
Stars with Giant Planets. ApJ, 622, 1075-1090, 2005.

\bibitem[Singer(2011)]{Singer}Singer, S. F.: Lack of consistency
between modeled and observed temperature trends. Energy \& Environ.
22 (4), 375\textendash{}406, 2011.

\bibitem[Solheim(2013a)]{Solheim}Solheim, J.-E.: The sunspot cycle
length \textendash{} modulated by planets?, Pattern Recogn. Phys.,
1, 159-164, doi:10.5194/prp-1-159-2013, 2013a.

\bibitem[Solheim(2013b)]{Solheim-b}Solheim, J.-E.: Signals from the
planets, via the Sun to the Earth, Pattern Recogn. Phys., 1, 177-184,
doi:10.5194/prp-1-177-2013, 2013b.

\bibitem[Spencer and Braswel(2011)]{Spencer}Spencer, R. W., Braswell,
W. D.: On the misdiagnosis of surface temperature feedbacks from variations
in earth's radiant energy balance. Remote Sens. 3, 1603\textendash{}1613,
2011.

\bibitem[Steinhilber and Beer(2013)]{Steinhilber} Steinhilber, F.,
Beer, J.: Prediction of solar activity for the next 500 years. Journal
of Geophysical Research: Space Physics 118, 1861\textendash{}1867,
2013.

\bibitem[Svensmark(2007)]{Svensmark}Svensmark, H.: Cosmoclimatology:
a new theory emerges. Astron. Geophys. 48, 18\textendash{}24, 2007.

\bibitem[Tan and Cheng(2012)]{Tan} Tan, B., Cheng, Z.: The mid-term
and long-term solar quasi-periodic cycles and the possible relationship
with planetary motions. Astrophys Space Sci. 343(2), 511-521, 2013.

\bibitem[Tattersall(2013)]{Tattersall}Tattersall, R.: The Hum: Lognormal
distribution of Planetary-Solar resonance. Pattern Recogn. Phys.,
1, 185-198, 2013.

\bibitem[Temple(1998)]{Temple} Temple, R.: The Sirius mystery. Appendix
IV ``Why Sixty Years?'', (Destiny Books, Rochester, Vermont) 1998.
\href{http://www.bibliotecapleyades.net/universo/siriusmystery/siriusmystery_appendix03.htm}{http://www.bibliotecapleyades.net/universo/siriusmystery/siriusmystery\_{}appendix03.htm}

\bibitem[Tinsley(2008)]{Tinsley}Tinsley, B. A.: The global atmospheric
electric circuit and its effects on cloud microphysics. Reports on
Progress in Physics 71, 066801, 2008.

\bibitem[Titius(1766)]{Titius}Titius, J. D.:, Betrachtung über die
Natur, vom Herrn Karl Bonnet (Leipzig, 1766), pp. 7-8; transl. by
Jaki, S., in ``The early history of the Titius-Bode Law'', American
Journal of Physics 40 1014-1023, 1972.

\bibitem[Tobias(2002)]{Tobias}Tobias, S. M.: The Solar Dynamo, Phil.
Trans. A, 360, 2741\textendash{}2756, 2002.

\bibitem[Vahrenholt and L\"uning(2013)]{Vahrenholt}Vahrenholt, F.,
L\"uning, S.: The neglected Sun, why the Sun precludes climate catastrophe.
(Stacey International, London), 2013. 

\bibitem[Voiculescu et al.(2013)]{Voiculescu}Voiculescu, M., Usoskin,
I., Condurache-Bota, S.: Clouds blown by the solar wind. Environ.
Res. Lett. 8, 045032, 2013.

\bibitem[Wang et al.(2012)]{Wang2012}Wang, Z., Wu, D., Song, X.,
Chen, X., Nicholls, S.: Sun\textendash{}moon gravitation-induced wave
characteristics and climate variation. J. Geophys. Res. 117, D07102,
2012.

\bibitem[Wilson(2011)]{Wilson2011}Wilson, I.R.G.: Are Changes in
the Earth\textquoteright{}s Rotation Rate Externally Driven and Do
They Affect Climate? The General Science Journal, 3811, 1-31, 2011.

\bibitem[Wilson(2013a)]{Wilson}Wilson, I. R. G.: The Venus\textendash{}Earth\textendash{}Jupiter
spin\textendash{}orbit coupling model, Pattern Recogn. Phys., 1, 147-158,
doi:10.5194/prp-1-147-2013, 2013a.

\bibitem[Wilson(2013b)]{Wilson2013b}Wilson, I. R. G.: Long-Term Lunar
Atmospheric Tides in the Southern Hemisphere, The Open Atmospheric
Science Journal, 7, 51-76, 2013b.

\bibitem[Wolf(1859)]{Wolf1859} Wolf, R.: Extract of a letter to Mr.
Carrington, Mon. Not. R. Astron. Soc., 19, 85\textendash{}86, 1859.

\bibitem[Wolff and Patrone(2010)]{Wolff} Wolff, C. L., Patrone, P.
N.: A new way that planets can affect the Sun. Solar Physics 266,
227-246, 2010.

\bibitem[Wyatt and Curry(2013)]{Wyatt}Wyatt, M. G., Curry, J. A.:
Role for Eurasian Arctic shelf sea ice in a secularly varying hemispheric
climate signal during the 20th century. Climate Dynamics. doi: 10.1007/s00382-013-1950-2,
2013 (in press). 

\bibitem[Zhou and Tung(2012)]{Zhou}Zhou, J., Tung, K.-K.: Deducing
multi-decadal anthropogenic global warming trends using multiple regression
analysis. J. Atmos. Sci., 70, 3\textendash{}8, doi:10.1175/JAS-D-12-0208.1.,
2012.\end{thebibliography}
\end{document}